\documentclass[%
 reprint,
 amsmath,
 amssymb,
 aps,
 prl]{revtex4-1}

\usepackage{diagbox}

\usepackage{graphicx}
\usepackage{dcolumn}
\usepackage{bm}
\usepackage[
  colorlinks=true,
  urlcolor=blue,
  linkcolor=blue,
  citecolor=blue
]{hyperref}
\usepackage{graphicx}
\usepackage{verbatim}
\usepackage{tabularx}
\usepackage{braket}
\usepackage{amsmath,amsfonts,amssymb}
\usepackage{graphicx}
\usepackage{float}
\usepackage{color}
\usepackage[verbose]{placeins}

\usepackage{fancyhdr}
\usepackage{accents}
\newlength{\dhatheight}
\newcommand{\doublehat}[1]{%
    \settoheight{\dhatheight}{\ensuremath{\hat{#1}}}%
    \addtolength{\dhatheight}{-0.35ex}%
    \hat{\vphantom{\rule{1pt}{\dhatheight}}%
    \smash{\hat{#1}}}}

\newcommand{\sket}[1]{\left|\left. #1 \right\rangle\!\right\rangle}
\newcommand{\sbra}[1]{\left\langle\!\left\langle #1 \right.\right|}
\newcommand{\sbraket}[2]{\left\langle\!\left\langle #1 | #2 \right\rangle\!\right\rangle}
\renewcommand{\vec}[1]{\text{vec}\left(#1\right)}
\newcommand{\tr}[1]{\mathrm{Tr}\left(#1\right)}

\begin{document}


\title{Demonstration of Universal Parametric Entangling Gates on a Multi-Qubit Lattice}

\author{M.~Reagor}\email{matt@rigetti.com}
\author{C.~B.~Osborn}
\author{N.~Tezak}
\author{A.~Staley}
\author{G.~Prawiroatmodjo}
\author{M.~Scheer}
\author{N.~Alidoust}
\author{E.~A.~Sete}
\author{N.~Didier}
\author{M.~P.~da~Silva}
\author{E.~Acala}
\author{J.~Angeles}
\author{A.~Bestwick}
\author{M.~Block}
\author{B.~Bloom}
\author{A.~Bradley}
\author{C.~Bui}
\author{S.~Caldwell}
\author{L.~Capelluto}
\author{R.~Chilcott}
\author{J.~Cordova}
\author{G.~Crossman}
\author{M.~Curtis}
\author{S.~Deshpande}
\author{T.~El~Bouayadi}
\author{D.~Girshovich}
\author{S.~Hong}
\author{A.~Hudson}
\author{P.~Karalekas}
\author{K.~Kuang}
\author{M.~Lenihan}
\author{R.~Manenti}
\author{T.~Manning}
\author{J.~Marshall}
\author{Y.~Mohan}
\author{W.~O'Brien}
\author{J.~Otterbach}
\author{A.~Papageorge}
\author{J.-P.~Paquette}
\author{M.~Pelstring}
\author{A.~Polloreno}
\author{V.~Rawat}
\author{C.~A.~Ryan}
\author{R.~Renzas}
\author{N.~Rubin}
\author{D.~Russell}
\author{M.~Rust}
\author{D.~Scarabelli}
\author{M.~Selvanayagam}
\author{R.~Sinclair}
\author{R.~Smith}
\author{M.~Suska}
\author{T.-W.~To}
\author{M.~Vahidpour}
\author{N.~Vodrahalli}
\author{T.~Whyland}
\author{K.~Yadav}
\author{W.~Zeng}
\author{C.~T.~Rigetti}

\affiliation{%
Rigetti Computing,
775 Heinz Avenue, Berkeley, CA 94710
}%

\date{\today}

\begin{abstract}
We show that parametric coupling techniques can be used to generate selective entangling interactions for multi-qubit processors. By inducing coherent population exchange between adjacent qubits under frequency modulation, we implement a universal gateset for a linear array of four superconducting qubits. An average process fidelity of $\mathcal{F}=93\%$ is estimated for three two-qubit gates via quantum process tomography. We establish the suitability of these techniques for computation by preparing a four-qubit maximally entangled state and comparing the estimated state fidelity against the expected performance of the individual entangling gates. In addition, we prepare an eight-qubit register in all possible bitstring permutations and monitor the fidelity of a two-qubit gate across one pair of these qubits. Across all such permutations, an average fidelity of $\mathcal{F}=91.6\pm2.6\%$ is observed. These results thus offer a path to a scalable architecture with high selectivity and low crosstalk.
\end{abstract}

\maketitle

\section{Introduction}

All practical quantum computing architectures must address the challenges of gate implementation at scale. Superconducting quantum processors designed with static circuit parameters can achieve high coherence times \cite{paik2011,Rigetti2012}. For these schemes, however, entangling gates have come at the expense of always-on qubit-qubit couplings \cite{Sheldon2016a} and frequency crowding \cite{Schutjens2013}. Processors based on tunable superconducting qubits, meanwhile, can achieve minimal residual coupling and fast multi-qubit operations \cite{Strauch2003, Barends2014}; yet, these systems must overcome flux noise decoherence \cite{Bialczak2007,Koch2007a} and computational basis leakage \cite{DiCarlo2009, Ghosh2013, Martinis2014, Biberio2017}. Moreover, the difficulties faced by both fixed-frequency and tunable qubit designs are compounded as the system size grows. Parametric architectures \cite{Bertet2006, niskanen2007quantum}, however, promise to overcome many of the fundamental challenges of scaling up quantum computers. By using modulation techniques akin to analog quantum processors  \cite{Flurin2015,Roy2016}, these schemes allow for frequency-selective entangling gates between otherwise static, weakly-interacting qubits.

Several proposals for parametric logic gates have been experimentally verified in the last decade. Parametric entangling gates have been demonstrated between two flux qubits via frequency modulation of an ancillary qubit \cite{Bertet2006,niskanen2007quantum}; between two transmon qubits via AC Stark modulation of the computational basis \cite{Majer2007} and of the non-computational basis \cite{Chow2013} with estimated gate fidelity of $\mathcal{F}\mathrm{=81}\%$ \cite{Chow2013}; between two fixed-frequency transmon qubits via frequency modulation of a tunable bus resonator with $\mathcal{F}\mathrm{=98}\%$ \cite{McKay2016}; between high quality factor resonators via frequency modulation of one tunable transmon \cite{Beaudoin2012,Strand2013,naik2017random} with $\mathcal{F}\mathrm{=[60-80]}\%$ \cite{naik2017random}; and finally, between a fixed-frequency and tunable transmon via frequency modulation of the same tunable transmon with $\mathcal{F}\mathrm{=93\%}$ \cite{TheoryPaper, BluePaper}. Despite these significant advances, there has yet to be an experimental assessment of the feasibility of parametric architectures with a multi-qubit system.

Here, we implement universal entangling gates via parametric control on a superconducting processor with eight qubits. We leverage the results of Refs.~\cite{TheoryPaper,BluePaper} to show how the multiple degrees of freedom for parametric drives can be used to resolve on-chip, multi-qubit frequency-crowding issues. For a four-qubit subarray of the processor, we compare the action of parametric CZ gates to the ideal CZ gate using quantum process tomography (QPT) \cite{Chuang1997, Poyatos1997, hradil20043}, estimating average gate fidelities \cite{Horodecki1999, Nielsen2002} of $\mathcal{F}=95\%$, $93\%$, and $91\%$. Next, we establish the scalability of parametric entanglement by comparing the performance of individual gates to the observed fidelity of a four-qubit maximally entangled state. Further, we directly quantify the effect of the remaining six qubits of the processor on the operation of a single two-qubit CZ gate. To do so, we prepare each of the 64 classical states of the ancilla qubit register and, for each preparation, conduct two-qubit QPT. Tracing out the measurement outcomes of the ancillae results in an average estimated fidelity of $\mathcal{F}=91.6\pm2.6\%$ to the ideal process of CZ. Our error analysis suggests that scaling to larger processors through parametric modulation is readily achievable.  
 
\begin{table*}[!htbp]
\setlength{\tabcolsep}{12pt}
\caption{\label{Table1} Characteristic parameters of the 8-qubit device. $\omega_r$ represents the frequency of the resonator, $\omega_{01}^{max}$ the qubit frequency (at zero flux), $\omega_{01}^{min}$ the frequency of the flux-tunable qubit at $\frac{1}{2}\Phi_0$, $\eta$ the anharmonicity of the qubit, $T_1$ the energy relaxation time of the qubit, $T_2^*$ the Ramsey phase coherence time, $\mathcal{F}_{RO}$ the single-shot readout assignment fidelity (* is a non-QND readout \cite{Reed2010}), and $p$ the single-qubit gate average error probability estimated as the decay of polarization under randomized benchmarking with Pauli generators of the Clifford group. Note that the anharmonicities of the flux-tunable qubits are measured at their operating frequencies.}
\centering
{\begin{tabular}{ccccccccc}
\hline
\hline
Qubit index & $\omega_r/2\pi$ & $\omega_{01}^{max}/2\pi$ & $\omega_{01}^{min}/2\pi$ & $-\eta/2\pi$ & $T_1$ & $T_2^*$ & $\mathcal{F}_{RO}$ & $p$ \\
 & (MHz) & (MHz) & (MHz) & (MHz) & ($\mu$s) & ($\mu$s) & (\%) & (\%) \\
\hline
$Q_0$ & 5065.0 & 3719.1 &    -   & 216.2 & 34.1 & 18.1 & 95.0 & 1.43 \\
$Q_1$ & 5278.0 & 4934.0 & 3817.9 & 204.0 &  17.0 &  4.3 & 93.2 & 0.70 \\
$Q_2$ & 5755.0 & 4685.8 &    -   & 199.4 & 14.2 & 12.9 & 93.7 & 1.02 \\
$Q_3$ & 5546.0 & 4870.9 & 3830.0 & 204.0 &  15.8 &  6.6 & 90.0 & 0.37 \\
$Q_4$ & 5164.0 & 4031.5 &    -   & 211.0 & 23.7 & 18.7 &  95.2*   & 0.70 \\
$Q_5$ & 5457.3 & 4817.6 & 3920.0 & 175.2 &  28.0 &  11.7 &  87.3*   & 2.00 \\
$Q_6$ & 5656.8 & 4662.5 &    -   & 196.6 & 16.9 & 15.4 &  93.8*   & 1.20 \\ 
$Q_7$ & 5388.1 & 4812.4 & 3803.5 & 182.8 &  5.6 &  8.6 &  89.9*   & 1.35 \\
\hline
\hline
\end{tabular}}
\end{table*}

\begin{figure}[!htbp]
\centering
\includegraphics[width=0.9\columnwidth]{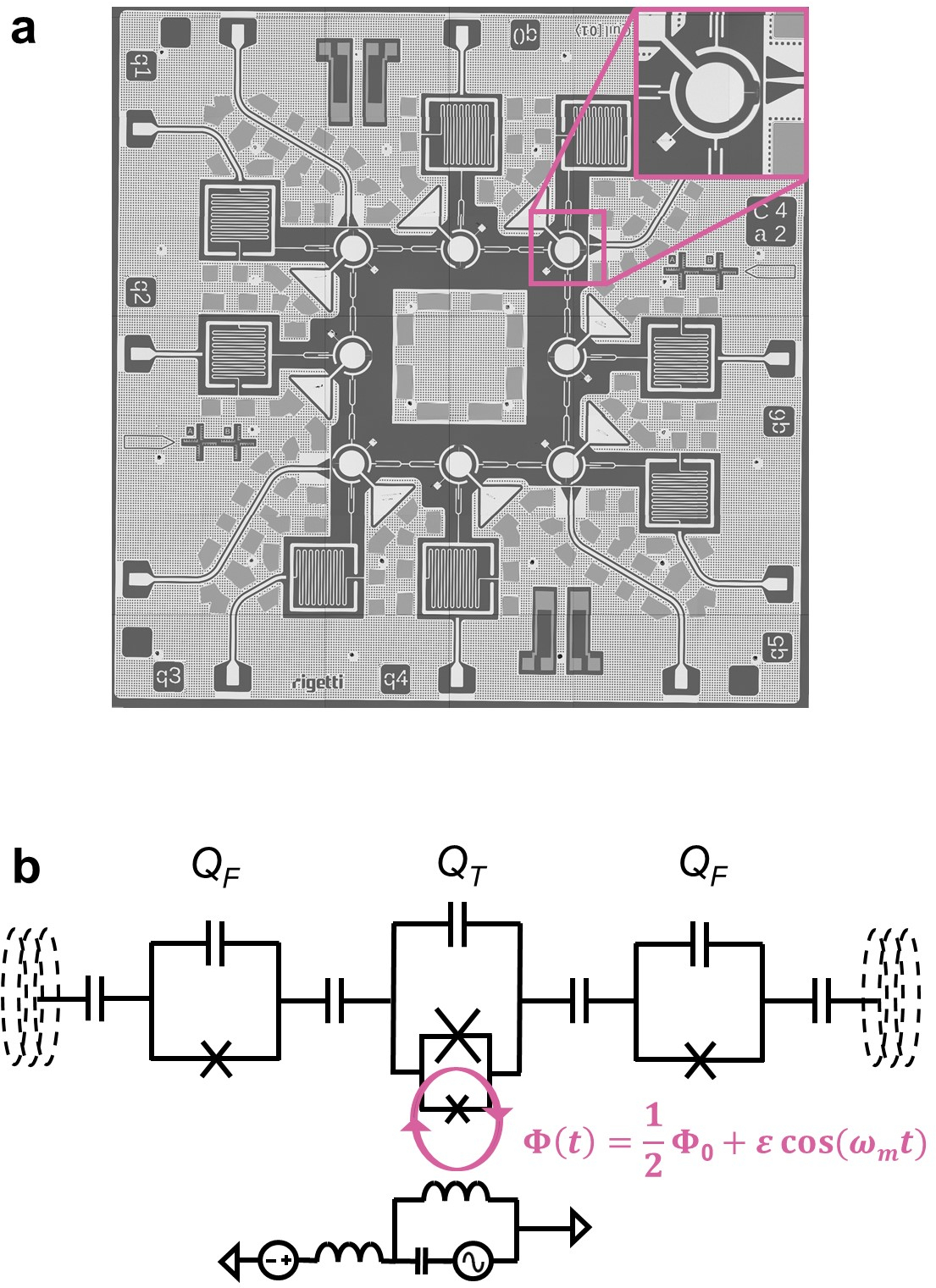}
\caption{\label{Fig-Device} \textbf{Device architecture.} \textbf{a,} Optical image of the 8-qubit superconducting circuit, consisting of 4 fixed-frequency ($Q_0$, $Q_2$, $Q_4$, $Q_6$) and 4 flux-tunable transmon qubits ($Q_1$, $Q_3$, $Q_5$, $Q_7$), used in the experiments. The inset shows a zoomed-in version of one of the tunable qubits. The dimensions of the chip are $5.5$ mm $\times$ $5.5$ mm. \textbf{b,} Circuit schematics of a chain of 3 qubits on the chip, where $Q_F$ represents the fixed transmons and $Q_T$ the tunable transmons. Each tunable qubit has a dedicated flux bias line connected to AC and DC drives combined using a bias tee, which tunes the time-dependent  magnetic flux $\Phi\text{(t)}$ threaded through its asymmetric SQUID loop, as depicted by the arrows.}
\end{figure}

\section{Device Design and Characteristics}

Figure~\ref{Fig-Device}a shows an optical image of the transmon qubit \cite{Koch2007} quantum processor used in our experiment. The multi-qubit lattice consists of alternating tunable and fixed-frequency transmons, each capacitively coupled to its two nearest neighbors to form a ring topology. This processor is fabricated on a high resistivity silicon wafer with 28 superconducting through-silicon vias (TSVs) \cite{FabPaper}. These TSVs improve electromagnetic isolation and suppression of substrate modes. Our fabrication process (See Ref.~\cite{FabPaper} and Supplementary Materials) requires deep reactive-ion etching (DRIE) and includes the deposition of  superconducting material into the etched cavity. A schematic of a triplet of transmons on the chip is shown in Fig.~\ref{Fig-Device}b, with a flux-delivery mechanism consisting of AC and DC drive sources, combined with a bias tee. The tunable transmons are designed with asymmetric Josephson junctions to provide a second flux-insensitive bias point \cite{Koch2007,Strand2013}. Characteristic parameters of all eight qubits are listed in Table \ref{Table1}. We observe an average energy relaxation time of $T_1=19.0$ $\mu$s and an average Ramsey phase coherence time of $T_2^*=12.0$ $\mu$s across the chip, despite the complexity of the fabrication process. We use randomized benchmarking \cite{Knill2008, Ryan2009, Chow2009} to estimate the average error probabilities of the single-qubit gates at an average of $p=1.1\%$, with the error estimated to be the decay constant of polarization for gates selected from the Pauli generators of the Clifford group. These coherence times and single-qubit gate fidelities allow us to accurately tomograph the parametric processes in this study.

Each qubit is coupled to an individual readout resonator for low crosstalk measurements. We operate in the dispersive regime \cite{Wallraff2004}, and use individual Josephson Parametric Amplifiers (JPAs) \cite{Roy2016} to amplify the readout signal. To calibrate the joint-qubit single-shot readout we iterate over all joint-qubit basis states, preparing each state 3000 times, and subsequently recording the time-averaged $I$ and $Q$ values of the returned signal for each qubit. By using a constant averaging filter over the demodulated returned signal, an average single-shot readout assignment fidelity of 92.3\% is achieved across the chip, as listed in Table \ref{Table1}. Using simultaneous multi-qubit readout, we train a separate binary classifier to predict the state of each qubit, accounting for readout crosstalk. The readout assignment fidelities quoted are defined as $\mathcal{F}_{RO}:= \frac{1}{2}[p(0|0)+p(1|1)]$ for each qubit.
Details on readout calibration are presented in the Supplementary Materials.

\section{Parametrically-activated entangling gates}

The Hamiltonian for a coupled tunable- and fixed-frequency transmon pair is well approximated by
\begin{align}
    \begin{split}
    \hat{H} / \hbar &=  \openone\otimes\left[
    \omega_T(t) \ket{1}\!\bra{1}
    + (2\omega_T(t) + \eta_T) \ket{2}\!\bra{2}\right]\\
    &+ \left[\omega_F \ket{1}\!\bra{1} 
    + (2\omega_F + \eta_F) \ket{2}\!\bra{2}\right]\otimes\openone\\
    &+ g \left(\sigma^{\dagger}_1 \otimes \sigma_2 + \mathrm{h.c.}\right),
    \end{split}
\end{align}
where $\omega_T$ ($\omega_F$) is the resonant frequency of the tunable- (fixed-) frequency transmon, $\eta_T$ ($\eta_F$) is
the corresponding anharmonicity, $g$ is the static capacitive
coupling between the transmons, and $\sigma_i=(\ket{0}\!\bra{1}+\sqrt{2}\ket{1}\!\bra{2})$. Modulating the flux through the SQUID loop sinusoidally results in
\begin{align}
\omega_T(t) = \overline{\omega}_T + \epsilon \cos(\omega_m t+\theta_m),
\end{align}
where $\omega_m$, $\epsilon$, and $\theta_m$ are the modulation frequency, amplitude, and phase, respectively,  $\overline{\omega}_T=\omega_T+\delta\omega$  is the average frequency and accounts for a time-independent frequency shift $\delta\omega$, leads to the interaction picture Hamiltonian
\cite{TheoryPaper,BluePaper}
\begin{align}
\label{eq1}
\hat{H}_\mathrm{int} / \hbar & = \sum_{n=-\infty}^\infty g_n\Big\{
e^{i(n\omega_m-\Delta)t}|10\rangle\langle01|\nonumber\\
&+\sqrt{2}e^{i(n\omega_m-[\Delta+\eta_F])t}|20\rangle\langle11|\nonumber\\
&+\sqrt{2}e^{i(n\omega_m-[\Delta-\eta_T])t}|11\rangle\langle02|\Big\}+\mathrm{h.c.},
\end{align}
where $g_n=g\,\mathrm{J}_n(\epsilon/\omega_m)e^{i\beta_n}$ are the
effective coupling strengths, $\Delta=\overline{\omega}_T-\omega_F$ is the effective detuning during modulation, 
$\beta_n=n(\theta_m+\pi)+\widetilde{\omega}_T\sin(\theta_m/\omega_m)$
is the interaction phase, and $\text{J}_n(x)$ are Bessel functions of the first kind.

\begin{figure}[!htbp]
\centering
\includegraphics[width=\columnwidth]{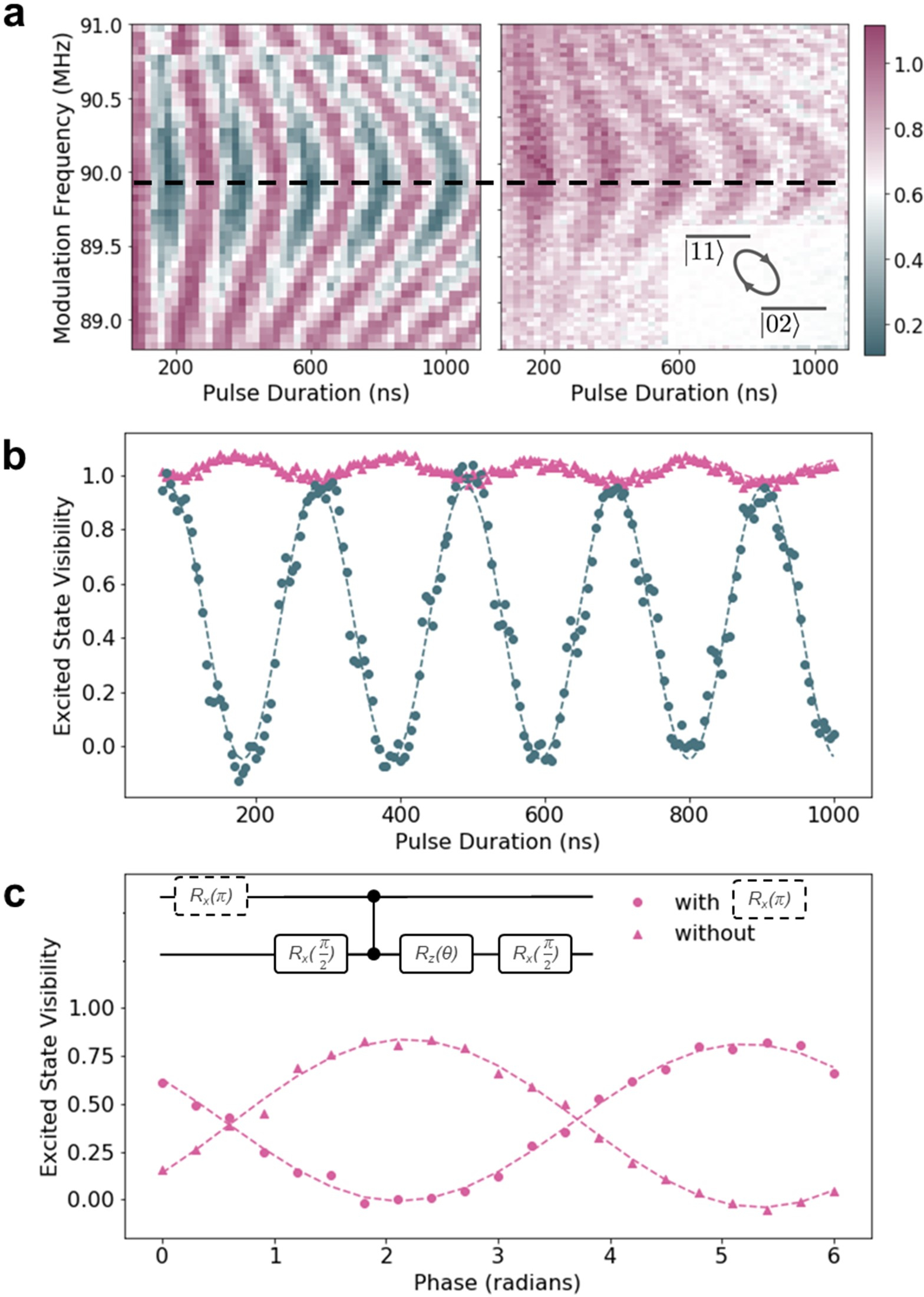}
\caption{\label{Fig-CZ} \textbf{Parametrically-activated entangling interactions.} \textbf{a,} (inset) Energy level diagrams of the $\ket{11}$ $\leftrightarrow$ $\ket{02}$ transition of $Q_0$ and $Q_1$. (main) Under modulation, coherent population exchange is observed within the $\ket{0}\leftrightarrow\ket{1}$ subspace of $Q_0$ (left), and within the $\ket{1}\leftrightarrow\ket{2}$ subspace for $Q_1$ (right). Excited state visibility axes are the averaged heterodyne signal of the readout pulse along an optimal $IQ$ quadrature axis, scaled to the separation in $IQ$ space of the attractors associated with ground and excited states of the qubits. \textbf{b,} Data from the dashed line in \textbf{a} shows the time-domain evolution between $Q_0$ and $Q_1$ on resonance, as teal (circles) and pink (triangles), respectively, allowing the identification of the target modulation duration of one period ($\tau=278$ns). \textbf{c,} (inset) Circuit diagram of the Ramsey interferometer to detect a geometric phase. (main) Determination of entangling-phase accumulation for the tunable qubit $Q_1$.}
\end{figure}

Parametric modulation of the tunable transmon's frequency is achieved by modulating the flux through the SQUID loop. As a result, $\overline{\omega}_T$ depends on the flux modulation amplitude, as well as the DC flux bias point~\cite{TheoryPaper}. Therefore, 
the resonance conditions for each of the terms in Eq.~\eqref{eq1} involve both the modulation 
amplitude and frequency. The first term in Eq.~\eqref{eq1} can be used to implement an iSWAP gate~\cite{SchuchPRA03,Strauch2003,BluePaper} of duration $\pi/2g_n$, while either of the latter two terms can be used to implement a CZ gate~\cite{DiCarlo2009,Ghosh2013,Chow2014,BluePaper} of duration $\pi/\sqrt{2}g_n$. In both cases, $n$ depends on the particular resonance condition. 
Although both gates are entangling and enable universal quantum computation when combined with single-qubit gates~\cite{Barenco,SchuchPRA03}, we choose to focus on the CZ implementation in order to reduce phase-locking constraints on room-temperature electronics. We thus calibrate three unique CZ gates: 
one between each of the neighboring pairs $(Q_0,Q_1)$, $(Q_1,Q_2)$, and $(Q_2,Q_3)$. 

\begin{table*}[!htbp]
\setlength{\tabcolsep}{12pt}
\caption{\label{Table2} Characteristics of the two-qubit CZ gates performed between neighboring qubit pairs ($Q_0, Q_1$), ($Q_1, Q_2$), and ($Q_2, Q_3$). $g_n$ represents the effective qubit-qubit coupling under modulation, $\omega_m$ is the qubit modulation frequency, $\delta\omega$ is the tunable qubit frequency shift under modulation, $\tau$ is the duration of the CZ gate, and $\mathcal{F}_{\rm QPT}$ is the two-qubit gate fidelity measured by quantum process tomography. The theoretical tunable qubit frequency shifts under modulation ($\delta\omega^{\rm th}/2\pi$) were obtained analytically using the experimentally determined modulation frequencies $\omega_m$ and are very close to the experimentally measured values ($\delta\omega/2\pi$). The gate durations and effective qubit-qubit couplings include pulse risetimes of  $40~\text{ns}$ to suppress the effect of pulse turn on phase.}
{\begin{tabular}{ccccccc}
\hline
\hline
Qubits & $g_n/2\pi$ & $(\omega_m/2)/2\pi$ & $\delta\omega^{\rm th}/2\pi$ & $\delta\omega/2\pi$ & $\tau$ & $\mathcal{F}_{\rm QPT}$ \\
 & (MHz) & (MHz) & (MHz) & (MHz) & (ns) & (\%) \\
\hline
$Q_0-Q_1$ & 2.53 &  83 & 270 & 281 & 278 & 95 \\
$Q_1-Q_2$ & 1.83 &  86 & 323 & 330 & 353 & 93 \\
$Q_2-Q_3$ & 1.59 & 200 & 257 & 257 & 395 & 91 \\
\hline
\hline
\end{tabular}}
\end{table*}

\begin{figure*}[!htbp]
\centering
\includegraphics[scale=0.75]{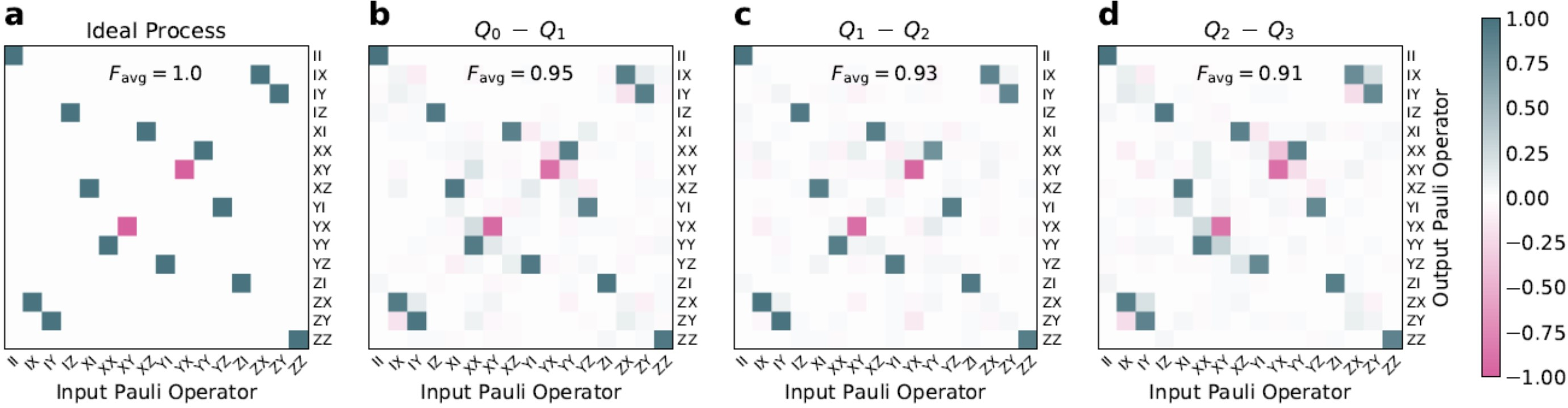}
\caption{\label{Fig-QPT} \textbf{Quantum process tomography.} Process matrices of \textbf{a,} the ideal process, and CZ gates between \textbf{b,} $Q_0-Q_1$, \textbf{c,} $Q_1-Q_2$, and \textbf{d,} $Q_2-Q_3$. The achieved average fidelities are measured to be 95\%, 93\%, and 91\%, respectively.}
\end{figure*}

The parametric CZ interaction between neighboring qubits can best be understood by examining the energy bands of the two-transmon subspace. Using the notation where $\ket{ij}$ corresponds to the $i$th energy level of the fixed-frequency qubit and the $j$th level of the tunable qubit, Fig.~\ref{Fig-CZ}a shows an example of the characteristic coherent oscillations that are produced as the modulation frequency of the tunable transmon is scanned through resonance with the $\ket{11}$ $\leftrightarrow$ $\ket{02}$ transition.

The CZ gate is activated by choosing modulation parameters that meet the resonance condition between $\ket{11}$ and $\ket{02}$ as implied in Eq.~\eqref{eq1}. This occurs when $\omega_m=(\overline{\omega}_T-\eta_T)-\omega_F$ and when the higher harmonics at $n \geq 2$ are also sufficiently detuned. Our device operates with a static flux bias of $\frac{1}{2} \Phi_0$, which makes the tunable qubit first-order insensitive to flux noise and modulation. The flux must be modulated, therefore, at a frequency of $\omega_m/2$ in order to meet the resonance condition for the gate~\cite{TheoryPaper}. This resonance condition results in an induced coherent population exchange between the $\ket{11}$ and $\ket{02}$ energy levels of the two-transmon subspace, shown for one pair of qubits in Fig.~\ref{Fig-CZ}a-b. After one cycle of oscillation in the population exchange between $\ket{11}$ and $\ket{02}$, all population returns to $\ket{11}$ (Fig.~\ref{Fig-CZ}c) with an additional geometric phase of $\pi$, achieving the desired CZ gate \cite{Strauch2003}.

The modulation parameters used in our parametric CZ gates are shown in Table \ref{Table2}. The modulation amplitude is a crucial tuning parameter for ensuring that a single interaction is activated during flux modulation, since the spectrum of induced coherent oscillations is a strong function of amplitude (see Supplementary Materials and Fig.~\ref{FigS-Birch}). We use the static frequency shift under modulation $\delta \omega$ to calibrate the effective drive amplitude in flux-quantum. The duration of the CZ gate $\tau$ is calibrated using measurements on coherent population exchange as shown in Fig.~\ref{Fig-CZ}b, with $\tau$ being one full period of the oscillation. In Fig.~\ref{Fig-CZ}c, two Ramsey measurements are performed on the tunable qubit; one with the fixed qubit in the $\ket{1}$ state, and the other with the fixed qubit in the $\ket{0}$ state. We remove the offset phase determined in this experiment by applying $R_Z(-\theta)$ in software at compilation time to the subsequent gates on the tunable qubit, which results in approximately the ideal CZ unitary of $\hat{U}=\textrm{diag}(1,1,1,-1)$.

\section{Process Tomography of parametric gates}

Next, we analyze our gates through quantum process tomography (QPT) \cite{Chuang1997, Poyatos1997, hradil20043}. Specifically, we characterize the behavior of each gate by reconstructing the evolution of a sufficiently large and diverse set of inputs, which corresponds to wrapping the gate by a set of pre- and post-rotations. We iterate over all pairs of rotations from the set $\hat R_j \in$ \{$\hat{\mathbb{I}}$, $\hat R_x$($\frac{\pi}{2}$), $\hat R_y$($\frac{\pi}{2}$), $\hat R_x$($\pi$)\} acting on each qubit separately. This yields a total of $16 \times 16 = 256$ different experiments, each of which we repeat $N=3000$ times. The single-shot readout data is classified into discrete positive-operator valued measure (POVM) outcomes. Assuming a multinomial model for each experiment, we can write the log-likelihood function for the full set of measurement records in terms of the histograms of POVM outcomes. This log-likelihood function is convex \cite{Boyd2004} in the quantum process matrix \cite{hradil20043}, allowing the use of the general purpose convex optimization package CVXPY~\cite{cvxpy} to directly solve the maximum likelihood estimation (MLE) problem (see Supplementary Materials for more details). 
Imposing complete positivity (CP) and trace preservation (TP) constraints on the estimated process is straightforward, as CVXPY also supports general semidefinite programs.
Using a basis of normalized multi-qubit Pauli operators (see Supplementary Materials) $\{\hat P_k,\; k=0,1,2,\dots,d^2-1\}$,
we represent a given process $\Lambda: \hat\rho \mapsto \Lambda(\hat\rho)$ in terms of the Pauli transfer matrix \cite{Chow2012} given by $(\mathcal{R}_\Lambda)_{kl}:= \mathrm{Tr} [\hat P_k \Lambda(\hat P_l)]$.

\begin{figure}[!htbp]
\centering
\includegraphics[width=\columnwidth]{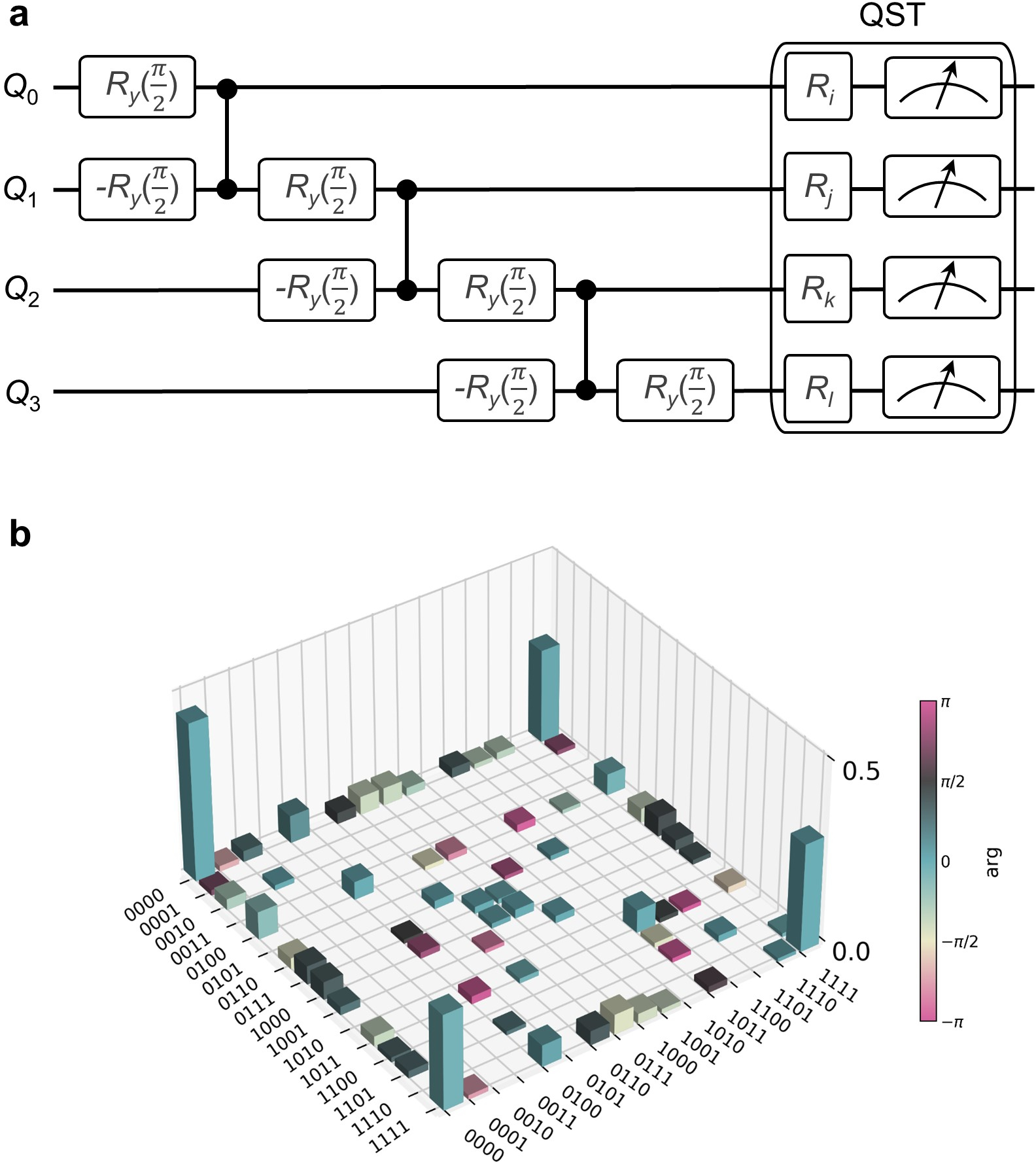}
\caption{\label{Fig-GHZ} \textbf{Quantum state tomography of GHZ state.} \textbf{a,} Quantum algorithm used to prepare the state $|\Psi\rangle=(|0000\rangle+|1111\rangle)/\sqrt{2}$ using CZ gates, and the quantum state tomography routine used to estimate the resulting density matrix. \textbf{b,} Reconstructed density matrix of the prepared GHZ estimated from quantum state tomography. The resulting state fidelity is estimated to be $\mathcal{F}=79\%$, in agreement with the expected performance of the three individual CZ gates, with color encoding the complex phase of each element. Density matrix elements below $|\rho_{\textrm{nm}}|\le0.01$ are cast transparent for visibility.}
\end{figure}

The Pauli transfer matrices obtained using the parametrically-activated CZ gates between $Q_0-Q_1$,  $Q_1-Q_2$, and  $Q_2-Q_3$ are shown in Fig.~\ref{Fig-QPT}b-d, with the ideal process matrix shown in Fig.~\ref{Fig-QPT}a. The average gate fidelity can be computed from the Pauli transfer matrix and is given by $\mathcal{F}= \frac{d^{-1}\mathrm{Tr} \mathcal{R}^T\mathcal{R}_{\rm CZ} + 1}{d+1}$, where $\mathcal{R}_{\rm CZ}$ is the Pauli transfer matrix of the ideal CZ gate.
The estimates obtained from process tomography for the average gate fidelity of the CZ operations between these pairs are $\mathcal{F}=95\%$, $93\%$, and $91\%$, respectively (Table~\ref{Table2} and Fig.~\ref{Fig-QPT}). To within less than 1\%, these results are confirmed when the MLE problem is solved under CP+TP physicality constraints.

\section{Infidelity analysis}

In this section, we analyze the contribution of seven potential error channels to the estimated average infidelity of a single CZ gate, between ($Q_1,Q_2$), with $1-\mathcal{F}=7\%$. For each potential error source, we establish an approximate upper bound contribution to the average infidelity. We use experiments to estimate five upper bounds and perform numerical simulations to estimate the others. A summary of these results can be found in Table~\ref{Table3}. We note that the sum of these bounds is greater than the estimated infidelity. We infer from this observation that some of these upper bounds are weak or that the effects of these errors do not combine linearly.

Decoherence mechanisms are the leading contributors to the infidelity of our gates. Operating the processor with tunable qubits statically biased to first-order insensitive flux bias points reduces the effect of flux noise on our gateset. However, coherence times are degraded during flux modulation due to the effective qubit frequency excursion from this first-order insensitive point. Furthermore, during flux modulation, the effective eigenvalues in the coupled subspace are a function of the modulation amplitude. Fluctuations in the modulation amplitude induce additional dephasing of the qubit. We measure the effective coherence time of the tunable qubits under modulation ($T^{*}_{2,\textrm{eff}}$), finding $T^{*}_{2,\textrm{eff}}~=~3-5.2~\mu$s during the parametric drives of the CZ gates (see Table~\ref{Table-Gate} in Supplementary Materials). These values are experimentally obtained by inserting a variable-time parametric drive into a Ramsey experiment. We estimate from comparing these times to the CZ gate durations that decoherence mechanisms of the tunable qubit should dominate the infidelity of our gates at the few percentage level for the calibrated gate durations.

To more precisely estimate the effect of decoherence on the fidelity of our two-qubit gates, we follow the procedure described in Eqs.~(11) and (12) in Ref.~\cite{BluePaper}. Specifically, we can estimate the unitary $\hat{V}$ that is nearest to our measured process $E$, and calculate its fidelity against the target unitary $\hat{U}$. The infidelity between $\hat{V}$  and $\hat{U}$ is entirely due to coherent errors (as both are coherent processes), and serves as a proxy for the coherent errors of $E$ with respect to $\hat{U}$. If $\hat{V}$  has high fidelity to $\hat{U}$, we take that to be an indication that the contribution from coherent errors is small. Additionally, if the infidelity between $E$ and $\hat{V}$  is similar to the infidelity between $E$ and $\hat{U}$, we take that to be an indication that the errors are dominated by decoherence. This is precisely the behavior we observe in our two-qubit gates, and is how we determine the contribution of decoherence to the average infidelity. 


We examine SPAM errors using a Maximum-Likelihood Estimation (MLE) method, which explicitly accounts for the nonideality of the readout by modeling it as a POVM, that we in turn estimate via separate readout calibration measurements. This implies that the readout infidelity is largely accounted for and corrected by our MLE tomography. The very large number of prepared bitstrings $(d=256)$ combined with the number of repetitions per preparation $(N=3000)$ results to a statistical uncertainty of $\approx1/\sqrt(3000 \times 256)\sim 0.1\%$. Even accounting for the qualitative nature of this argument, we expect the error due to an imperfectly estimated readout model to be significantly smaller than 1\%.

Errors in single-qubit gates will affect the observed infidelity of a QPT experiment because these are used for pre- and post-rotations. To account for their contributions, we independently measure the infidelity of the tomography pre- and post-rotation gates via simultaneous randomized benchmarking (SRB) experiments, which are based on sequences that  uniformly sample the non-entangling subgroup of the two-qubit Clifford group. A rough estimate based on the gate duration and decoherence time yields an expected infidelity of $\sim0.5-1\%$. The SRB experiments confirm this and yield a typical infidelity of $\sim1\%$ for our tomographic pre- and post-rotations. An estimate for the resulting upper bound on the infidelity of the CZ process matrix would thus be $\sim1-2\%$, as there is a separate pre- and post-rotation for each QPT measurement sequence.

Because of weak anharmonicity of transmon qubits, leakage  to  the  non-computational  subspace  also  con-tributes to the infidelity of the entangling gates. We bound  leakage error by preparing the two qubits in $|11\rangle$ and applying the parametric gate. In doing so, we exit and enter the computational subspace. Imprecise control of this operation results to residual population in the transmon's second excited state. We measure this residual population in the $|2\rangle$ state to be 6\% after the QPT measurement is completed. Because population out of the computational basis is unaffected by the QPT post-rotations, this population behaves as an extra decoherence channel. Bounding the resulting infidelity to the CZ gate as the full population is a worst-case approximation.


Undesired changes in the amplitude or frequency of the modulation pulse (due to instrument imperfections, temperature variations, etc.), moreover, result in an unwanted shift of the qubit effective frequency under modulation, $\bar \omega_{T}$, introducing infidelity to a QPT experiment. The effect of the former is straightforward, but the latter is a combined result of the amplitude-frequency interdependence of this modulation technique, and of the frequency-dependent signal transfer function through the system. This leads us to calculate
\begin{equation}
\mathrm{d} F = \frac{\partial F}{\partial \bar\omega_{T}} \frac{\partial \bar\omega_{T}}{\partial t}  \mathrm{d}t.
\end{equation}
We estimate $\partial \bar\omega_{T}/\partial t$ from measurements taken over several hours, during which we see worst-case excursions in $\bar\omega_{T}$ of roughly 1 MHz per hour. For a full process tomography measurement, $t \sim 5$ min, resulting in a maximum frequency excursion of $\partial \bar\omega_{T}/\partial t \cdot \partial t \simeq  0.08$ MHz. We estimate $\partial F/\partial \bar\omega_{T}$ from the linewidth of the gate's chevron pattern (Fig.~\ref{Fig-CZ}a), which ranges between 2 and 4 MHz. Assuming a linear loss in fidelity for shifts away from the gate's frequency (chevron's center frequency), we calculate $\partial F/\partial \bar\omega_{T} \simeq 1 / (2~ \text{MHz})$. Hence, $\mathrm{d} F = (\partial F/\partial \bar\omega_{T})(\partial \bar\omega_{T}/\partial t) \mathrm{d} t\simeq 0.08~\text{MHz} / 2~\text{MHz} \simeq 0.04$, which provides an estimate of the contribution of undesired changes in the  modulation  pulse to the average infidelity. 

Moreover, we quantitatively estimate the last two sources of error (i.e. spurious sidebands and residual ZZ coupling) using theoretical simulations after measuring the spectrum and qubit-qubit $\chi$. The Hamiltonian expressed in the interaction picture, Eq.~\eqref{eq1}, is composed of two kinds of coupling: the always-on capacitive couplings, not specifically activated by the modulation, which correspond to the terms with $n=0$. and spurious sidebands that correspond to $n\neq0$. To estimate the effect of always-on coupling and spurious sidebands on the gate fidelity, we simulate the system with the relevant coupling terms separately, and estimate their contributions to be $\sim1.9\%$ and $\sim0.03\%$, respectively.

\begin{table}[!tbp]
\setlength{\tabcolsep}{12pt}
\caption{\label{Table3} Error analysis for the two-qubit CZ gate between pairs ($Q_1,Q_2$). Contributions to the average infidelity estimated from QPT for several error channels.}
{\begin{tabular}{lc}
\hline
\hline
Error channel & Contribution to average \\
or process & infidelity bound ($\le$\%) \\
\hline
Decoherence & 6.5 \\
SPAM error & 0.2 \\
Tomography rotations & 2.0 \\
Leakage into $|02 \rangle$ & 6.0 \\
Residual ZZ coupling & 1.9 \\
Spurious sidebands & 0.03 \\
Instrumentation drift & 1.0 \\
\hline
\hline
\end{tabular}}
\end{table}

\section{Four-qubit GHZ state}

We benchmark the multi-qubit action of these parametric gates by running a quantum algorithm (Fig.~\ref{Fig-GHZ}) that ideally prepares a maximally entangled, four-qubit GHZ state \cite{Barends2014}, followed by the execution of quantum state tomography (QST) \cite{Roos2004,Steffen2006} on the resulting four-qubit state. The same set of tomography post-rotations used for QPT are also used here for QST. Similar convex optimization techniques to QPT (see Supplementary Materials for more details) allow the tomographic inversion required to estimate the density matrix for QST. The reconstructed density matrix is shown in Fig.~\ref{Fig-GHZ}. We compute a resulting state fidelity, $\mathcal{F}=\langle\Psi|\hat \rho|\Psi\rangle$, to an ideal four-qubit GHZ state, $|\Psi\rangle=(|0000\rangle+|1111\rangle)/\sqrt{2}$, of $\mathcal{F}=79\%$. This holds both with and without the positivity $\hat \rho \ge 0$ constraint applied in the estimation. Assigning all of the resulting state error to the action of the CZ gates results in an estimate for a geometric mean of $\mathcal{F}=92\%$ for the three two-qubit gates, which is a difference of $0.5\%$ from the geometric mean estimated from individual QPT analysis. We therefore conclude that further improvements to the fidelity of individual two-qubit operations will translate to improved algorithmic fidelities on this multi-qubit lattice.  

\begin{figure}[!htbp]
\centering
\includegraphics[width=\columnwidth]{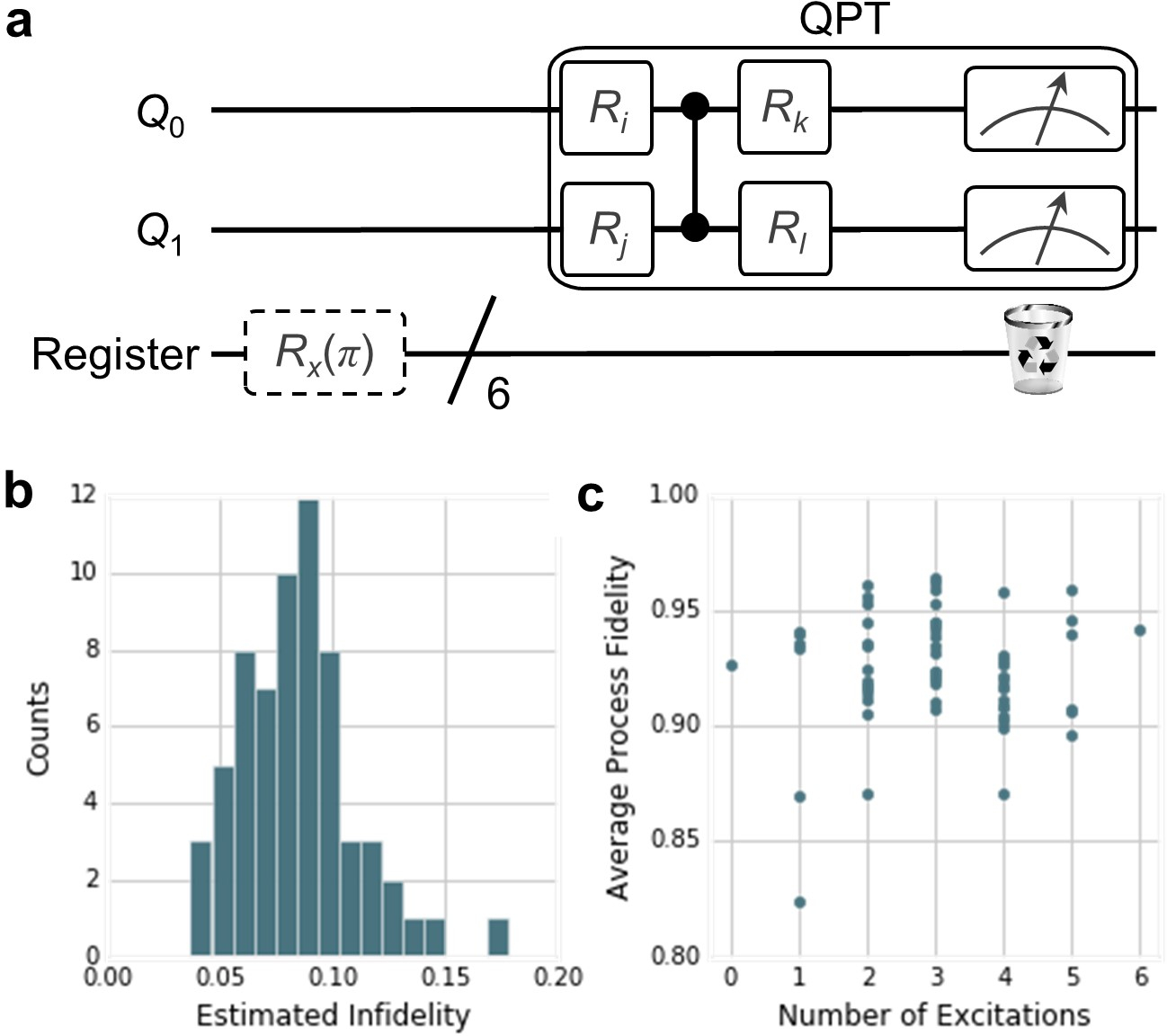}
\caption{\label{Fig-6Q-QPT} \textbf{Crosstalk.} \textbf{a,} Pulse sequences used for quantifying the effect of crosstalk from ancilla qubits on the performance of CZ gates. To do this, first an arbitrary bitstring register of six ancilla qubits is prepared, with each qubit in either the ground or excited state. Then, process tomography is performed on the CZ gate between the other two qubits on the 8-qubit chip, to extract a fidelity.  \textbf{b,} Histogram of the estimated infidelities measured using this algorithm. \textbf{c,} Average process fidelities achieved as a function of the number of excited qubits in the ancilla register.}
\end{figure}

\section{Quantifying Crosstalk}

In order to detect the coherent errors that are introduced by the effects of residual qubit-qubit coupling (with those qubits not associated with a certain two-qubit gate), we run a tomography procedure that involves all eight qubits of the processor. The circuit diagram for this measurement is shown in Fig.~\ref{Fig-6Q-QPT}a. After first preparing all qubits in the ground state, we apply single-qubit rotations on a sub-register of six ancilla qubits ($Q_2-Q_7$); applying either the identity gate or $R_X(\pi)$ to these qubits for a given run. Immediately thereafter, we run QPT for a CZ gate between the remaining pair of qubits ($Q_0-Q_1$). We repeat this procedure 64 times, once for each unique bitstring of the six qubit register. For signal-to-noise considerations, each bitstring experiment is performed 250 times. The total experiment thus amounts to 4.1$\times10^{6}$ individual measurements. The histogram of the estimated infidelities is shown in Fig.~\ref{Fig-6Q-QPT}b. 
While the mean of the distribution is $\mathcal{F}=91.6\pm2.6\%$, there are a few outliers with infidelities that are larger by a statistically significant amount. Surprisingly, the worst estimated gate performance is observed for bitstrings in which one of the next-nearest-neighbor qubits ($Q_3$) is excited (rather than a nearest-neighbor of the pair). We attribute this error to the dispersive interaction between $Q_3$ and both $Q_0$ and $Q_1$: we measure these dispersive shifts to be $\delta \omega_{0,3}/2\pi$~=~150kHz and $\delta \omega_{1,3}/2\pi$~=~270kHz. For a CZ gate duration of $\tau$~=~278~ns, these shifts correspond to single qubit phase accumulation of approximately $\delta\theta_0$=0.26~rad and $\delta\theta_1$=0.47~rad, which we associate with the observed drop in QPT fidelity. Increasing the static detuning between $Q_1$ and $Q_3$ in future designs, which is 14.5~MHz here, is expected to reduce this error channel by the squared ratio of the new detuning to the current detuning.

In addition, the estimated process fidelity versus the number of excited ancilla qubits for all measured bitsrings of the register is shown in Fig.~\ref{Fig-6Q-QPT}c. Despite the observed variations, the average process fidelities for all but three bitstrings are within the standard error of the experiment. This demonstrates that the two-qubit parametrically-activated CZ gate is mostly insensitive to the ancilla qubits, compared to architectures that must directly address qubit-qubit coupling effects. This is a critical property of scalable quantum processors. It is worth noting, however, that the worst-case gate error estimates for such entangling gate should be considered for purposes such as error correction schemes.

With no need for intermediary couplers, we have demonstrated a parametric scheme for performing universal quantum computation on a four-qubit subarray of an eight-qubit processor. By doing so, we have reduced circuit design complexity and simplified the procedure to generate multi-qubit entangling gates, in a manner that is both frequency selective and alleviates the challenges of frequency crowding. We have measured two-qubit gate fidelities up to 95\% on the subarray, and demonstrated limited sensitivity of these gates to the state of an ancilla register of the remaining six qubits. Ongoing work with this processor includes the demonstration of eight-qubit algorithms, as well as further benchmarking via multi-qubit randomized benchmarking \cite{Ryan2009,Gaebler2012,Corcoles2013} and gate-set tomography~\cite{Blume-Kohout2013, Greenbaum2015, Blume-Kohout2017}. Our results also highlight improvable parameters for future devices that utilize this architecture, which provides a promising foundation for high-fidelity, scalable quantum processors.

\section{Acknowledgements} Part of this work was performed at the Stanford Nano Shared Facilities, supported by the NSF under award ECCS-1542152. \textbf{Funding:} This work was funded by Rigetti \& Co Inc., dba Rigetti Computing. \textbf{Author contributions}: M. Reagor, C.B.O., N.T., A.S., G.P., M. Scheer, and N.A. designed and performed the experiments and analyzed the data. E.A.S., N.D., and M.P.d.S. proposed and performed theoretical analysis of the experimental results. M. Reagor, N.A., C.B.O., N.T., A.S., P.K., E.A.S., N.D., and M.P.d.S. performed figure planning and prepared the manuscript draft. M. Reagor and C.T.R. were responsible for the overall direction, planning, and integration among different research units. E.A., J.A., A. Bestwick, M.B., B.B., A. Bradley, C.B., S.C., L.C., R.C., J.C., G.C., M.C., S.D., T.E.B., D.G., S.H., A.H., P.K., K.K., M.L., R.M., T.M., J.M., Y.M., W.O., J.O., A. Papageorge, J.-P.P., M.P., A. Polloreno, V.R., C.A.R., R.R., N.R., D.R., M. Rust, D.S., M. Selvanayagam, R. Sinclair, R. Smith, M. Suska, T.-W.T., M.V., N.V., T.W., K.Y., and W.Z. provided hardware and software support for performing the experiments, designing and fabricating the chip, and theoretically analyzing the results.
\textbf{Competing interests:} C.T.R. is the founder and chief executive officer of Rigetti \& Co Inc. All authors are, have been, or may in the future be participants in incentive stock plans at Rigetti \& Co Inc. E.A.S., N.D., M.P.d.S., C.T.R., M. Reagor, S.C., N.T., and C.A.R. are inventors on two pending patent
applications related to this work (no. 62/521,943, filed 19 June 2017; no. 62/573,446, filed 17
October 2017). The other authors declare that they have no competing interests. \textbf{Data and materials
availability:} All data needed to evaluate the conclusions in the paper are present in the paper and/or the Supplementary Materials. Additional data related to this paper may be requested from the authors.

\section{References}

\bibliography{RQC_4Q}

\begin{thebibliography}{10}

\bibitem{paik2011}
H.~Paik, {\it et~al.\/}, {\it Phys. Rev. Lett.\/} {\bf 107}, 240501 (2011).

\bibitem{Rigetti2012}
C.~Rigetti, {\it et~al.\/}, {\it Phys. Rev. B\/} {\bf 86}, 100506(R) (2012).

\bibitem{Sheldon2016a}
S.~Sheldon, E.~Magesan, J.~M. Chow, J.~M. Gambetta, {\it Phys. Rev. A\/} {\bf
  93}, 060302 (2016).

\bibitem{Schutjens2013}
R.~Schutjens, F.~A. Dagga, D.~J. Egger, F.~K. Wilhelm, {\it Phys. Rev. A\/}
  {\bf 88}, 052330 (2013).

\bibitem{Strauch2003}
F.~W. Strauch, {\it et~al.\/}, {\it Phys. Rev. Lett.\/} {\bf 91}, 167005
  (2003).

\bibitem{Barends2014}
R.~Barends, {\it et~al.\/}, {\it Nature\/} {\bf 508}, 500 (2014).

\bibitem{Bialczak2007}
R.~C. Bialczak, {\it et~al.\/}, {\it Phys. Rev. Lett.\/} {\bf 99}, 187006
  (2007).

\bibitem{Koch2007a}
R.~H. Koch, D.~P. DiVincenzo, J.~Clarke, {\it Phys. Rev. Lett.\/} {\bf 98},
  267003 (2007).

\bibitem{DiCarlo2009}
L.~DiCarlo, {\it et~al.\/}, {\it Nature\/} {\bf 460}, 240 (2009).

\bibitem{Ghosh2013}
J.~Ghosh, A.~G. Fowler, J.~M. Martinis, M.~R. Geller, {\it Phys. Rev. A\/} {\bf
  88}, 062329 (2013).

\bibitem{Martinis2014}
J.~M. Martinis, M.~R. Geller, {\it Phys. Rev. A\/} {\bf 90}, 022307 (2014).

\bibitem{Biberio2017}
H.~Ribeiro, A.~Baksic, A.~A. Clerk, {\it Phys. Rev. X\/} {\bf 7}, 011021
  (2017).

\bibitem{Bertet2006}
P.~Bertet, C.~J. P.~M. Harmans, J.~E. Mooij, {\it Phys. Rev. B\/} {\bf 73},
  064512 (2006).

\bibitem{niskanen2007quantum}
A.~O. Niskanen, {\it et~al.\/}, {\it Science\/} {\bf 316}, 723 (2007).

\bibitem{Flurin2015}
E.~Flurin, N.~Roch, J.~D. Pillet, F.~Mallet, B.~Huard, {\it Phys. Rev. Lett.\/}
  {\bf 114}, 090503 (2015).

\bibitem{Roy2016}
A.~Roy, M.~Devoret, {\it C. R. Phys.\/} {\bf 17}, 740  (2016).

\bibitem{Majer2007}
J.~Majer, {\it et~al.\/}, {\it Nature\/} {\bf 449}, 443 (2007).

\bibitem{Chow2013}
J.~M. Chow, {\it et~al.\/}, {\it New J. Phys.\/} {\bf 15}, 115012 (2013).

\bibitem{McKay2016}
D.~C. McKay, {\it et~al.\/}, {\it Phys. Rev. Applied\/} {\bf 6}, 064007 (2016).

\bibitem{Beaudoin2012}
F.~Beaudoin, M.~P. {Da Silva}, Z.~Dutton, A.~Blais, {\it Phys. Rev. A\/} {\bf
  86}, 022305 (2012).

\bibitem{Strand2013}
J.~D. Strand, {\it et~al.\/}, {\it Phys. Rev. B\/} {\bf 87}, 220505 (2013).

\bibitem{naik2017random}
R.~K. Naik, {\it et~al.\/}, {\it arXiv preprint arXiv:1705.00579\/}  (2017).

\bibitem{TheoryPaper}
N.~Didier, E.~A. Sete, M.~P. da~Silva, C.~T. Rigetti, {\it arXiv preprint
  arXiv:1706.06566\/}  (2017).

\bibitem{BluePaper}
S.~Caldwell, {\it et~al.\/}, {\it arXiv preprint arXiv:1706.06562\/}  (2017).

\bibitem{Chuang1997}
I.~L. Chuang, M.~A. Nielsen, {\it J. Mod. Opt.\/} {\bf 44}, 2455 (1997).

\bibitem{Poyatos1997}
J.~F. Poyatos, J.~I. Cirac, P.~Zoller, {\it Phys. Rev. Lett.\/} {\bf 78}, 390
  (1997).

\bibitem{hradil20043}
Z.~Hradil, J.~{\v{R}}eh{\'a}{\v{c}}ek, J.~Fiur{\'a}{\v{s}}ek, M.~Je{\v{z}}ek,
  {\it Quantum state estimation\/}, M.~Paris, J.~\v{R}eh\'{a}\v{c}ek, eds.
  (Springer-Verlag, Berlin, Heidelberg, 2004), pp. 59--112.

\bibitem{Horodecki1999}
M.~Horodecki, P.~Horodecki, R.~Horodecki, {\it Phys. Rev. A\/} {\bf 60}, 1888
  (1999).

\bibitem{Nielsen2002}
M.~A. Nielsen, {\it Phys. Lett. A\/} {\bf 303}, 249  (2002).

\bibitem{Reed2010}
M.~D. Reed, {\it et~al.\/}, {\it Phys. Rev. Lett.\/} {\bf 105}, 173601 (2010).

\bibitem{Koch2007}
J.~Koch, {\it et~al.\/}, {\it Phys. Rev. A\/} {\bf 76}, 042319 (2007).

\bibitem{FabPaper}
M.~Vahidpour, {\it et~al.\/}, {\it arXiv preprint arXiv:1708.02226\/}  (2017).

\bibitem{Knill2008}
E.~Knill, {\it et~al.\/}, {\it Phys. Rev. A\/} {\bf 77}, 012307 (2008).

\bibitem{Ryan2009}
C.~A. Ryan, M.~Laforest, R.~Laflamme, {\it New J. Phys.\/} {\bf 11}, 013034
  (2009).

\bibitem{Chow2009}
J.~M. Chow, {\it et~al.\/}, {\it Phys. Rev. Lett.\/} {\bf 102}, 090502 (2009).

\bibitem{Wallraff2004}
A.~Wallraff, {\it et~al.\/}, {\it Nature\/} {\bf 431}, 162 (2004).

\bibitem{SchuchPRA03}
N.~Schuch, J.~Siewert, {\it Phys. Rev. A\/} {\bf 67}, 032301 (2003).

\bibitem{Chow2014}
J.~M. Chow, {\it et~al.\/}, {\it Nat. Commun.\/} {\bf 5}, 4015 (2014).

\bibitem{Barenco}
A.~Barenco, {\it et~al.\/}, {\it Phys. Rev. A\/} {\bf 52}, 3457 (1995).

\bibitem{Boyd2004}
S.~S. Boyd, L.~Vandenberghe, {\it {Convex optimization}\/} (Cambridge
  University Press, Cambridge, 2004).

\bibitem{cvxpy}
S.~Diamond, S.~Boyd, {\it J. Mach. Learn. Res.\/} {\bf 17}, 1 (2016).

\bibitem{Chow2012}
J.~M. Chow, {\it et~al.\/}, {\it Phys. Rev. Lett.\/} {\bf 109}, 060501 (2012).

\bibitem{Roos2004}
C.~F. Roos, {\it et~al.\/}, {\it Phys. Rev. Lett.\/} {\bf 92}, 220402 (2004).

\bibitem{Steffen2006}
M.~Steffen, {\it et~al.\/}, {\it Science\/} {\bf 313}, 1423 (2006).

\bibitem{Gaebler2012}
J.~P. Gaebler, {\it et~al.\/}, {\it Phys. Rev. Lett.\/} {\bf 108}, 260503
  (2012).

\bibitem{Corcoles2013}
A.~D. C{\'{o}}rcoles, {\it et~al.\/}, {\it Phys. Rev. A\/} {\bf 87}, 030301(R)
  (2013).

\bibitem{Blume-Kohout2013}
R.~Blume-Kohout, {\it et~al.\/}, {\it arXiv preprint arXiv:1310.4492\/}
  (2013).

\bibitem{Greenbaum2015}
D.~Greenbaum, {\it arXiv preprint arXiv:1509.02921\/}  (2015).

\bibitem{Blume-Kohout2017}
R.~Blume-Kohout, {\it et~al.\/}, {\it Nat. Commun.\/} {\bf 8}, 14485 (2017).

\bibitem{scikit-learn}
F.~Pedregosa, {\it et~al.\/}, {\it J. Mach. Learn. Res.\/} {\bf 12}, 2825
  (2011).

\bibitem{Tibshirani-ISL}
G.~James, D.~Witten, T.~Hastie, R.~Tibshirani (Springer-Verlag, New York,
  2013).

\bibitem{Nielsen2010}
M.~A. Nielsen, I.~L. Chuang, {\it {Quantum Computation and Quantum
  Information}\/} (Cambridge University Press, Cambridge, 2010).

\bibitem{Gelman-BDA}
A.~Gelman, {\it et~al.\/}, {\it Bayesian Data Analysis\/} (CRC Press, 2013).

\bibitem{Jeffrey2014}
E.~Jeffrey, {\it et~al.\/}, {\it Phys. Rev. Lett.\/} {\bf 112}, 1 (2014).

\bibitem{Magesan2014}
E.~Magesan, J.~M. Gambetta, A.~D. C{\'{o}}rcoles, J.~M. Chow, {\it Phys. Rev.
  Lett.\/} {\bf 114}, 200501 (2014).

\end{thebibliography}
\bibliographystyle{Science}





\pagebreak
\widetext

\begin{titlepage}
	\centering
	{\large \textbf{Demonstration of Universal Parametric Entangling Gates on a Multi-Qubit Lattice} \\ \normalfont Supplementary Materials \par}

\end{titlepage}
\setcounter{equation}{0}
\setcounter{figure}{0}
\setcounter{table}{0}
\setcounter{page}{1}
\makeatletter
\renewcommand{\theequation}{S\arabic{equation}}
\renewcommand{\thefigure}{S\arabic{figure}}
\renewcommand{\thetable}{S\Roman{table}}
\renewcommand{\bibnumfmt}[1]{[S#1]}
\renewcommand{\citenumfont}[1]{S#1}


\subsection{Fabrication and Design}

The eight-qubit device is fabricated on a high-resistivity ($>$ 15 k$\Omega$cm) intrinsic silicon wafer with Rigetti through-silicon vias (TSVs). The process is explained in more detail in \cite{FabPaper}. The backside of the wafer is blanket Al film while the device side has two metal layers deposited on it. A layer of 85nm/5nm/60nm Al/Ti/Pd serves as an alignment mark for Josephson junction (JJ) electron beam lithography (EBL) followed by a 160nm Al device layer. The Al pattern includes readout resonators with coupling arms to the qubits. Both layers are patterned with optical lithography and metal deposited with e-beam evaporation techniques followed by an NMP/IPA lift-off process. A 400V, 20mA, 5$^{\circ}$ argon ion-milling process is used to remove the native oxide of the Al film in the vias prior to device-side Al deposition. This provides a high conducting connection between the metal in the vias and the device side.
 
\begin{figure*}[!htbp]
\centering
\includegraphics[scale=0.75]{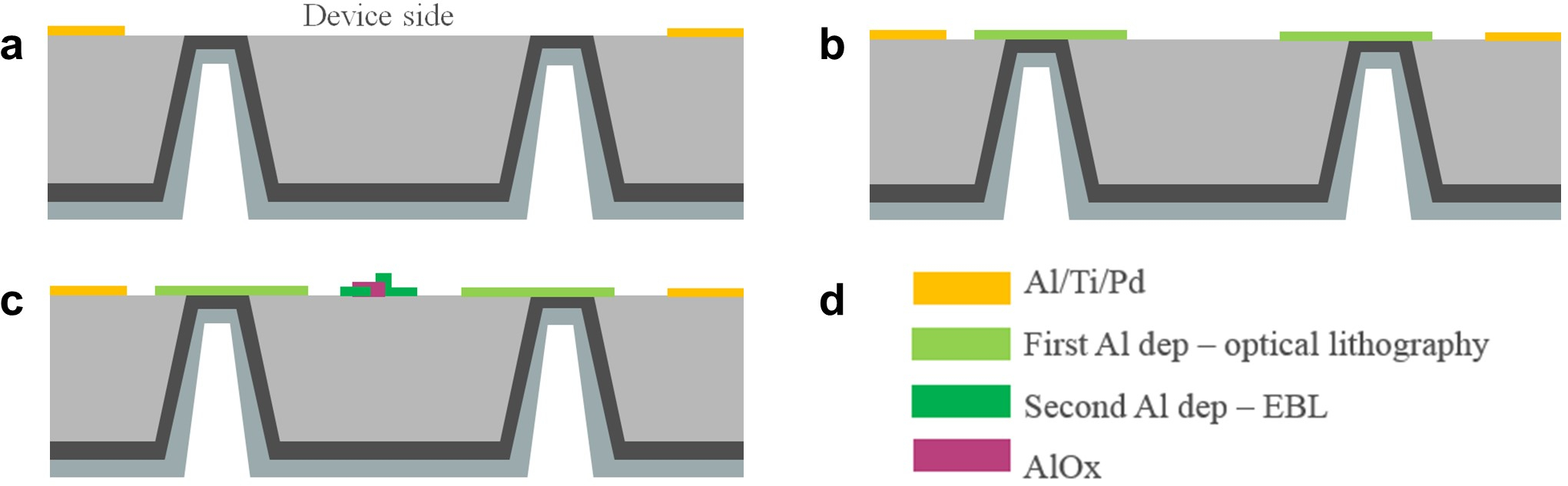}
\caption{\label{FigS-Fab} Steps of the fabrication process of the 8-qubit quantum processor.}
\end{figure*}

Next, the wafer is cleaved into dies and goes through transmon qubit patterning. First the dies are cleaned with acetone, IPA, and UV descum and then coated with bilayer MMA/PMMA resist stack. Then, EBL techniques are used to pattern the transmon capacitances and JJs. The die is placed into a high-vacuum electron-beam evaporation chamber and after a gentle ion-milling step (200 V, 8mA, 20$^{\circ}$), a double-angle evaporation technique at 20$^{\circ}$ is used to deposit Al/AlOx/Al layer. Finally, hot NMP followed by IPA is used to lift off the film. The fabrication process steps are presented in Fig.~\ref{FigS-Fab}.

\subsection{Theoretical predictions of the gate parameters}

\begin{figure*}[!htbp]
\centering
\includegraphics[scale=0.75]{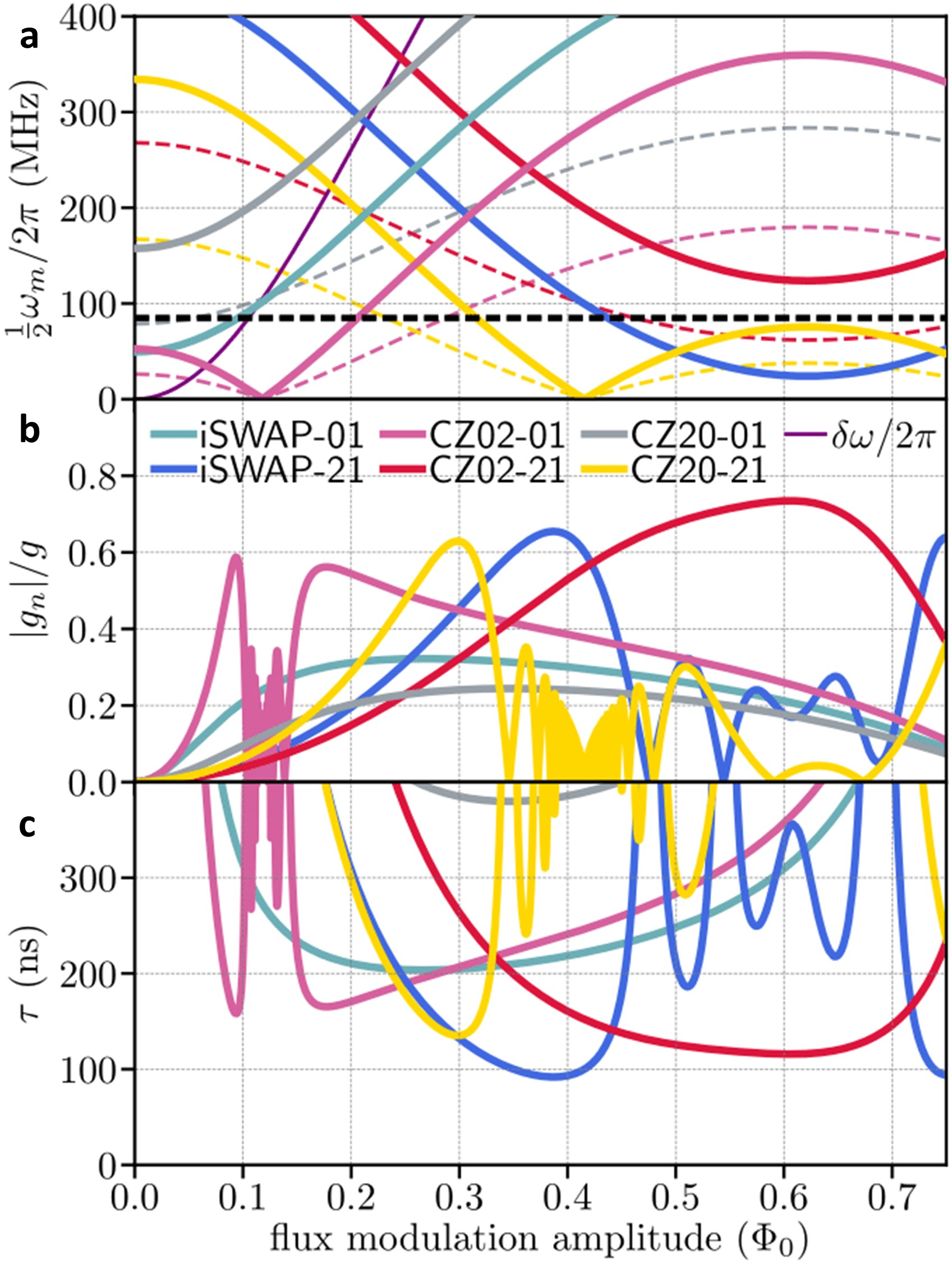}
\caption{\label{FigS-Birch} Theoretical predictions of \textbf{a}, modulation frequency, \textbf{b}, effective coupling under modulation, and \textbf{c}, gate time as a function of the flux pulse modulation amplitude for activating a given gate (iSWAP, CZ$_{02}$, CZ$_{20}$) between the pairs $(Q_0,Q_1)$ and $(Q_1,Q_2)$.}
\end{figure*}

Figure \ref{FigS-Birch} shows the theoretical parameters of parametric two-qubit gates as a function of the pulse modulation amplitude for the pairs $(Q_0,Q_1)$ and $(Q_1,Q_2)$. The modulation frequency required to activate a given gate (iSWAP, CZ$_{02}$, CZ$_{20}$) is plotted in Fig.~\ref{FigS-Birch}a. The full line of each gate is the modulation frequency, $\omega_m/2$, that activates the first harmonics. The dashed line is the modulation frequency for the second harmonics equal to $\omega_m/4$. An efficient entangling interaction takes place at a modulation amplitude where the modulation frequency is away from spurious sidebands. The frequency shift $\delta\omega$ of the tunable qubit is plotted in purple. The effective coupling under modulation is plotted in Fig.~\ref{FigS-Birch}b for the first harmonics (divided by the bare coupling). The gate time is plotted in Fig.~\ref{FigS-Birch}c. It corresponds to the entangling interaction time for a flat pulse. All parameters are calculated analytically from the results of Ref.~\cite{TheoryPaper}. For completeness, we have reported an additional set of experimental and theoretical parameters of the two-qubit CZ gates in Table~\ref{Table-Gate}.

\begin{table*}[!htbp]
\setlength{\tabcolsep}{8pt}
\caption{\label{Table-Gate} Characteristics of the two-qubit CZ gates performed between neighboring qubit pairs ($Q_0, Q_1$), ($Q_1, Q_2$), and ($Q_2, Q_3$). $\Delta_{11\leftrightarrow02}$ represents the detuning between $\ket{11}$ and $\ket{02}$, $\Delta_{11\leftrightarrow20}$ the detuning between $\ket{11}$ and $\ket{20}$, $T^{*}_{2,\textrm{eff}}$ the effective coherence time of the tunable qubit under modulation, $n$ the associated harmonics, and $\phi_{p}$ the modulation amplitude in units of flux quantum. The symbol $^{\dagger}$ denotes the transitions used for the gate.}
\centering
{\begin{tabular}{ccccccc}
\hline
\hline
Qubit pair & $\Delta_{11\leftrightarrow02}/2\pi$ & $\Delta_{11\leftrightarrow20}/2\pi$ & $T^{*}_{2,\textrm{eff}}$ & Harmonics & $\phi_{p}$ \\
index & (MHz) & (MHz) & ($\mu$s) & $n$ & ($\Phi_{0}$) \\
\hline
$Q_0-Q_1$ & 69.2$^{\dagger}$ & -315.0 & 3.8  & -1 & 0.20 \\
$Q_1-Q_2$ & 1035.9 & 668.5$^{\dagger}$ & 3.0 & 2 & 0.23 \\
$Q_2-Q_3$ & 1041.6 & 656.4$^{\dagger}$ & 5.2 & 1 & 0.21 \\
\hline
\hline
\end{tabular}}
\end{table*}


\subsection{Single-Shot Readout}

\begin{figure*}[!htbp]
\centering
\includegraphics[scale=.5]{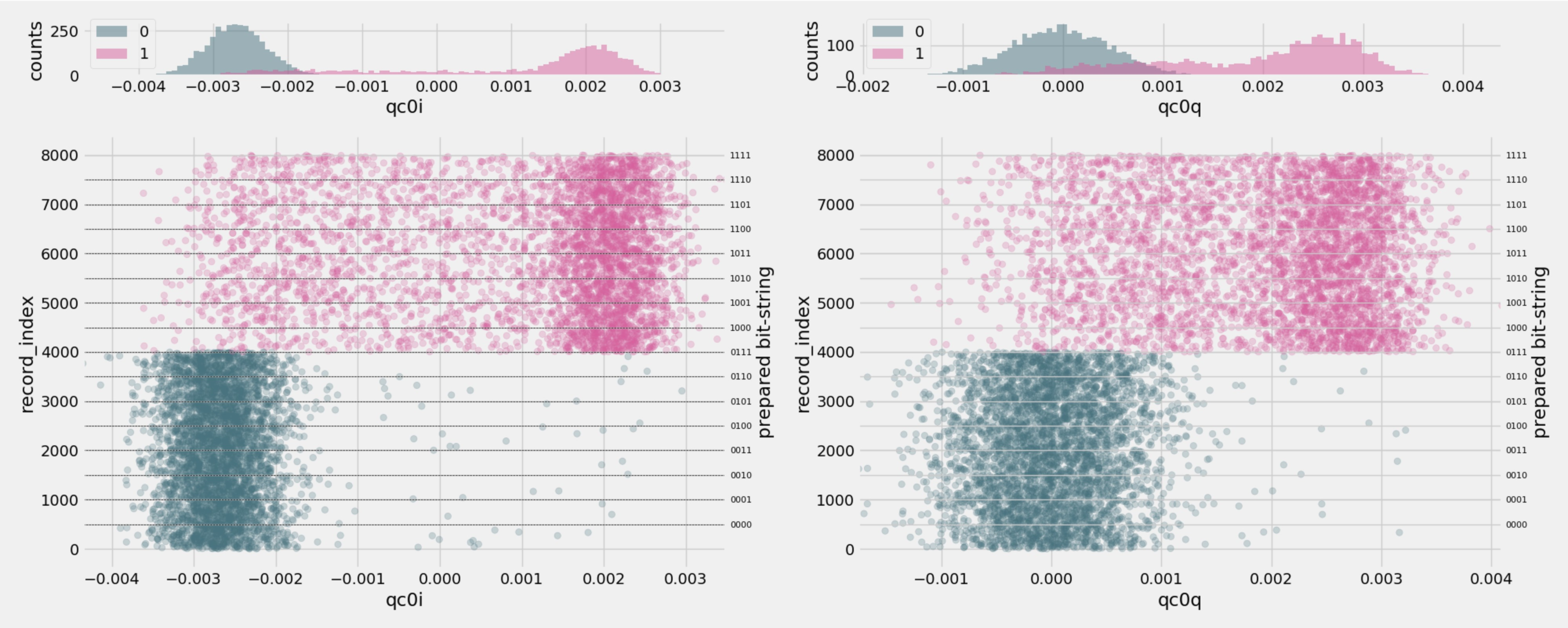}
\caption{\label{FigS1} To detect systematic errors in the full readout of the prepared bitstring it is helpful to visualize the raw data as a scatter plot against the prepared bitstring. The above figure shows the $I, Q$ readout data of qubit 0, colored according to its prepared state. Hence the color oscillates with a period given by the position of the qubit in the bitstring.
We aggregate the different prepared states over all bitstrings into a global histogram. As explained in more detail in the main text, good separation between the states of an individual qubit directly translates to a good classifier.}
\end{figure*}
The states of our quantum processor are read out by interrogating the resonator for each qubit with flat pulse of roughly $1\mu{\rm s}$ duration that is near resonant with the linear resonator and can detect the qubit state based on a dispersive frequency shift \cite{Wallraff2004}. The signals are amplified to increase the signal to noise ratio relative to the thermal noise that is added in the return path back to room temperature. Each sampled readout signal is integrated against a simple boxcar filter, yielding a scalar complex value per qubit $\overline{z} = I + i Q = \int_{t_0}^{t_0+T}z(t)dt$.
To infer the underlying discrete qubit states from the noisy continuous data, we need to classify the readout data and assign a state.

To this end we implement automated classifiers through four separate three-fold cross-validated L2-penalized logistic regressions, trained on a stratified subset of the measured data. The decision thresholds for the individual classifiers are determined with the remaining holdout set by maximizing the Kolmogorov-Smirnov (KS) statistic. The classifier's performance may be characterized via its confusion rate matrix $p_{jk} := p(\text{predicted } j \mid\text{ prepared }k)$, which can also be used to write an effective positive operator valued measure (POVM) for single-shot measurement including classification $\hat{N}_j = \sum_{k=0}^{d-1} p(\text{predicted } j \mid\text{ prepared }k)\hat \Pi_k$, where $\hat \Pi_k:=\ket{k}\bra{k}$ are projectors onto the joint qubit states $k=0, 1, \dots , d-1$.

The quality of the simultaneous single-shot readout data of the 4-qubit chain ($Q_\text{0}-Q_\text{3}$) may be visualized using scatter plots of (stratified samples of) the measured data, as shown in Fig.~\ref{FigS1}. We sort the scatter according to the prepared bitstrings and color each point according to the prepared state of the individual qubit whose $(I, Q)$ scatter is shown. Hence the frequency of color change is correlated with the position of the qubit in the bitstring. This enables us to visually detect possible readout cross-talk, where the readout of one qubit depends on the state of another. The histograms at the top of each panel show the qubit readout data aggregated over all prepared bitstrings. As before, the color indicates the prepared state of the qubit. A good separation between histograms of different color minimizes the overlap and thus misclassification. This property is directly connected to the ability to train a powerful classifier, as will be discussed in more detail in the next section.

\subsection{N-Qubit Readout Linear Classifier}

\begin{figure*}[!htbp]
\centering
\includegraphics[scale=0.55]{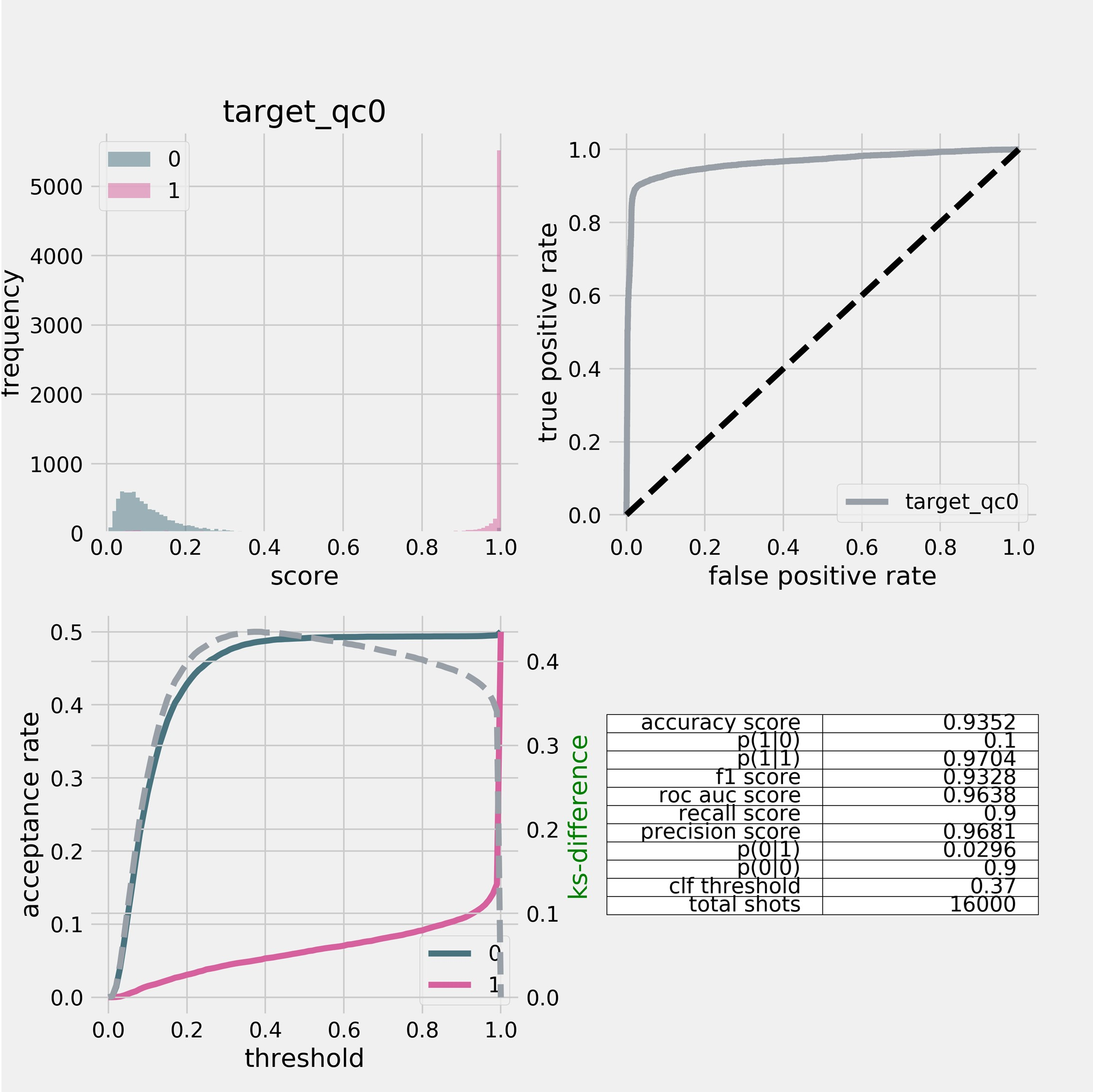}
\caption{\label{fig:classifier_perf} We evaluate the performance of the individual trained classifier for qubit 0's state with different metrics shown in the above figure. The top left corner shows the score distribution of the classifier when evaluated on a holdout set. A good separation is key to a good classifier, and the asymmetry is an indicator that the original data is not completely independent and Gaussian. The top right plot shows the Receiver-Operator-Characteristic (ROC) curve. The lower left plot shows the Kolmogorov-Smirnov (KS) statistic that is being used to infer the optimal readout threshold. Finally, the lower right table lists different Figures-of-Merit, such as the confusion rates, the precision, recall, etc. For more detail please refer to the main text of this Supplementary Materials.}
\end{figure*}

To optimize readout fidelities we take into account the fact that the readout signal for a given qubit $i$ is not fully independent of the state of the other qubits ($j\neq i$), due to non-negligible cross-talk between different qubits on the chip. For $n=4$ qubits, each with an $(I, Q)$ readout pair, we have a combined $p=8$ dimensional feature vector $\mathbf{f_r}$ for each single shot  $r=1,\dots,M$. To train the classifier we first split the data into $m_t$ training and $m_h$ holdout records, $M=m_t+m_h$, using stratified sampling based on the prepared bitstring that we use as target variable $\mathbf{y_r}=(y_{r,q})_{q=1,...n}$ for the classifier. The stratification is done using \textit{scikit-learn's (v0.18.1)} \texttt{StratifiedShuffleSplit}  \cite{scikit-learn} with the random seed set to $42$ and a holdout set fraction of $20\%$. We proceed by pre-processing the training data before fitting the final classifiers: (1) we standardize the data to have zero mean and unit standard deviation (per feature in the feature vector) and (2) decorrelate the feature space using a Principal Component Analysis (PCA) \cite{Tibshirani-ISL}. Both methods are readily available in \textit{scikit-learn}. It is worthwhile to point out that the standardization is a natural transformation for our data as single-shot noise has Gaussian character \cite{Nielsen2010}. After the pre-processing is finished, we split the target variable into $n=4$ individual targets $y_r\in\{0,1\}$ corresponding to the prepared states of each individual qubit at shot $r$. We use those targets to train $n=4$ individual $L2$-penalized logistic regressions using the full $p=8$ dimensional pre-processed training feature vector for each qubit, i.e. we take the (pre-processed) data of each of the other qubits into account when fitting the target qubit classifier (this also means that in order to classify the readout of a qubit, measuring all the other qubits is needed to ensure obtaining all features). The training is done using \texttt{LogisticRegressionCV} which automates a three-fold stratified cross-validation procedure to find an optimal value of the $L2$-penalty coefficient based on maximizing the accuracy score of the classifier with a default decision threshold of $0.5$. Here, the accuracy score for an individual classifier is defined as
\begin{equation}
    \text{acc} = 1-\frac{1}{m_t} \sum_{i=1}^{m_t} |y_i - \hat y_i| \in [0,1]
\end{equation}
where the predicted class is
\begin{equation}
    \hat y_i = \begin{cases}
        1, \text{ if } p(y_i=1|\mathbf{f_i}) \geq 0.5\\
        0 \text{ else}.
        \end{cases}
\end{equation}
The score $p_i = p(y_i=1|f_i)$ is given using the logistic link function
\begin{equation}
    \text{logit}\,p_i = \text{log}\frac{p_i}{1-p_i} = -\beta_0-\pmb\beta_1\cdot\mathbf{f_i} + \lambda |\pmb\beta_1|^2.
    \label{eq:logit}
\end{equation}
Here the values $\beta_0$, $\pmb\beta_1$, and $\lambda$ are the intercept, the weight-vector and the $L2$-penalty, respectively. All of these parameters are inferred during the training of the classifier. It should be noted that, despite looking like a probability, $p_i$ is not calibrated due to the presence of the penalty $\lambda$ and hence, represents only a score in the range $[0,1]$. The presence of the penalty can be motivated by observing that the noise model is not purely given by Gaussian noise, but in general can also contain noise from other sources, such as thermal qubit population, $T_1$-decay, and non-linear readout effects. This necessitates the introduction of a regularizer to avoid the impact of potential high-leverage points in the fitting procedure and reduce overfitting of the classifier to the training set \cite{Gelman-BDA}. Note that the $L2$-penalty can be understood as a Gaussian prior on the weights $\beta_0$ and $\pmb\beta_1$. To see this let us assume a purely Gaussian noise model
\begin{equation}
    y_n = \pmb\beta \mathbf{x}_n + \epsilon
\end{equation}
where $\epsilon$ is normally distributed with zero mean and variance $\sigma^2$, i.e. $\epsilon \sim \mathrm{N}(0,\sigma^2)$. The Gaussian likelihood is
\begin{equation}
    L = \Pi_{n=1}^N \mathcal{N}(y_n|\pmb\beta\mathbf{x}_n,\sigma^2).
\end{equation}
Now, if some additional information about $\pmb\beta$ is given according to a Gaussian distribution this prior can be incorporated using Bayes rule to obtain
\begin{equation}
    L = \Pi_{n=1}^N \mathcal{N}(y_n|\pmb\beta\mathbf{x}_n,\sigma^2)\mathcal{N}(\pmb\beta|0,\lambda^{-1}).
\end{equation}
Taking the logarithm of the likelihood $L$ we obtain Eq.~\ref{eq:logit} up to some additive and multiplicative constants, that are irrelevant in the optimization routine.

After fitting the classifier, i.e.,~inferring the values $\beta_0$, $\pmb\beta_1$, and $\lambda$, we use the holdout set containing $m_h$ previously unseen readout samples to evaluate the classifiers. The results are shown in Figs.~\ref{fig:classifier_perf}; each individual figure evaluates the performance of the individual qubit classifiers. The upper left inset shows a histogram over the scores for all $m_h$ shots, where the color code is representing the true, i.e. prepared, state of the qubit. Strong separation between the clusters indicates that reliable classification can be done. The upper right corner shows the Receiver-Operator-Characteristic (ROC) curve \cite{Tibshirani-ISL}, which illustrates the diagnostic ability of the classifier as the threshold is varied. For a given threshold $t$ we calculate the fraction of correctly identified positive labels (true positive rate $tpr$) and  falsely identified positive labels (false positive rate $fpr$) and plot those against each other (See below for a more rigorous definition of $tpr$ and $fpr$). In an ideal scenario we would identify all positive labels without ever encountering a falsely assigned label, which would make the curve approach the upper left corner. If we assigned labels uniformly at random, we would obtain the dashed diagonal line. The further away the ROC curve is from the dashed line the better the classifier is.

As we already pointed out above the score $p_i$ is not a true calibrated probability and hence we should infer an optimal decision threshold $t^*$ to maximize the performance of the classifier. A useful metric to determine $t^*$ is to maximize the Kolmogorov-Smirnov (KS) statistic. We define the two non-parametric distributions over the positive (1) and negative (0) classes individually
\begin{align}
    F_\text{pos}(t) =&\frac{1}{n_\text{pos}}\sum_{r=1}^{n_\text{pos}} I_{[0, t]}(p_i(y_i=1)),\\
    F_\text{neg}(t) =&\frac{1}{n_\text{neg}}\sum_{r=1}^{n_\text{neg}} I_{[0, t]}(p_i(y_i=0)),
\end{align}
where we refer to the different bit classes as positive and negative, $m_h= n_\text{pos}+n_\text{neg}$ and $I_{[0, t]}$ being the indicator function on the interval $[0, t]$. Then we can determine the optimal threshold $t^*$ as
\begin{equation}
    t^* = \underset{t\in[0,1]}{\text{max}} \left|F_\text{pos}(t) -  F_\text{neg}(t) \right|.
\end{equation}
The curves for $F_\text{pos}(t)$, $F_\text{neg}(t)$, and their difference are shown in the lower left plots of Figs.~\ref{fig:classifier_perf} and clearly show that the optimal thresholds are widely varying for different qubit classifiers.

Finally the table in the lower right corner contains several summary metrics \cite{Tibshirani-ISL}. Before introducing any metrics let us define the concept of a confusion matrix. In any (binary) classification problem we have the prepared (true) labels and the inferred (predicted) labels. We can hence introduce four quantities in a matrix
\begin{center}
    \begin{tabular}{|c|c|c|}
      \hline
      \diagbox[innerwidth=1.5cm]{true}{pred} & 0 & 1\\
      \hline
      0 & $TN$ & $FP$ \\
      \hline
      1 & $FN$ & $TP$ \\
      \hline
    \end{tabular}
\end{center}
Here TN are true negatives, TP are true positives, FN are false negatives, and FP are false positives. The sum of $TN+TP+FN+FP = M$ where M is the total number of records. It is straightforward to generalize this to the multi-label classification case. Note that the counts of these quantities depend strongly on the threshold that is applied to the classification score to decide on a positive or negative label. We can normalize these quantities and obtain conditional misclassification probabilities $p(\text{predicted}\; j \mid \text{prepared}\; k)$ as $p(0|0) = \frac{TN}{TN + FP}$, $p(1|0)  = \frac{FP}{TN + FP}$, $p(0|1) = \frac{FN}{FN + TP}$ and $p(1|1) = \frac{TP}{FN + TP}$. We can now define several interesting quantities:
\begin{itemize}
    \item The accuracy score $\text{acc}=\frac{TP + TN}{M}$ tells us how well the classifier identifies overall correct samples.
    \item The true positive rate (a.k.a. recall) $tpr = \frac{TP}{TP + FN} = p(1|1)$ gives us an estimate of how sensitive the classifier is for a correct identification of a positive sample given all positive cases.
    \item The false positive rate $fpr = \frac{FP}{FP+TN} = p(1|0)$ which gives us an estimate of how often we wrongly assign a postive label given a negative true label.
    \item The precision $prec=\frac{TP}{TP + FP}$ is an estimate for the rate of correctly classifying a positive label given the assignment of a positive label.
    \item The $F_1$-score $F_1=2\frac{tpr \cdot prec}{tpr + prec}$ gives us an estimate of how well balanced the classifier is in its trade-off between precision and recall.
    \item The ROC AUC, short for Receiver-Operator-Characteristic Area-Under-Curve, is the integral of the ROC curve
        \begin{align}
            \begin{split}
            \text{AUC} &= \int dt -\text{tpr}(t)\frac{d}{dt}\text{fpr}(t) \\
            &= P(p_i(y_i=1) >p_j(y_j=0))
            \end{split}
        \end{align}
    and can be interpreted as the probability that the score $p_i$ of a randomly chosen positive sample $i$ is higher than the score $p_j$ of a randomly chosen negative sample $j$.
\end{itemize}

Note that we generally do not need to use the joint N-qubit readout to accurately predict a given qubit's state. We do occasionally observe readout crosstalk, but this is usually restricted to direct neighbors. For predicting a given qubit's state it is therefore sufficient to use only its readout and that of its direct neighbors. This implies that we need not prepare all $d=2^n$ measurement basis states to train our readout classifiers. A much smaller number that is strictly less than $2^e n$ is sufficient, where $e$ is the maximum number of neighbors of any qubit in the lattice.
Finally, we expect that even the small readout crosstalk we currently observe will be greatly reduced by ongoing improvements to our quantum processor design and packaging.

\subsection{Quantum Process Tomography}

In the following we use `super-ket' notation $\sket{\rho} := \vec{\hat\rho}$ where $\vec{\hat\rho}$ is a density operator $\hat\rho$ collapsed to a single vector by stacking its columns. The standard basis in this space is given by $\{\sket{j},\; j=0,1,2\dots, d^2-1\}$, where $j=f(k,l)$ is a multi-index enumerating the elements of a $d$-dimensional matrix row-wise, i.e. $j=0 \Leftrightarrow (k,l)=(0,0)$, $j=1 \Leftrightarrow (k,l)=(0,1)$, etc. The super-ket $\sket{j}$ then corresponds to the operator $\ket{k}\bra{l}$.

We similarly define $\sbra{\rho} := \vec{\hat\rho}^\dagger$ such that the inner product $\sbraket{\chi}{\rho} = \vec{\hat\chi}^\dagger \vec{\hat\rho} = \sum_{j,k=0}^{d^2-1} \chi_{jk}^\ast\rho_{jk} = \tr{\hat{\chi}^\dagger \hat\rho}$ equals the Hilbert-Schmidt inner product. If $\rho$ is a physical density matrix and $\hat{\chi}$ a Hermitian observable, this also equals its expectation value. When a state is represented as a super-ket, we can represent super-operators acting on them as $\Lambda \to \doublehat{\Lambda}$, i.e., we write $\sket{\Lambda(\hat\rho)} = \doublehat{\Lambda}\sket{\rho}$.

We introduce an orthonormal, Hermitian basis for a single qubit in terms of the Pauli operators and the identity $\sket{ P_j} := \vec{\hat P_j}$ for $j=0,1,2,3$, where $\hat P_0 = \mathbb{\hat I}/\sqrt{2}$ and $\hat P_k=\sigma_{k}/\sqrt{2}$ for $k=1,2,3$. These satisfy $\sbraket{ P_l}{ P_m}=\delta_{lm}$ for $l,m=0,1,2,3$. For multi-qubit states, the generalization to a tensor-product basis representation carries over straightforwardly. The normalization $1/\sqrt{2}$ is generalized to $1/\sqrt{d}$ for a
d-dimensional space. In the following we assume no particular size of the system.

We can then express both states and observables in terms of linear combinations of Pauli-basis super-kets and super-bras, respectively, and they will have real valued coefficients due to the hermiticity of the Pauli operator basis. Starting from an initial state $\rho$ we can apply a completely positive map to it
\begin{align}
       \hat \rho' = \Lambda_K(\hat\rho) =  \sum_{j=1}^n \hat K_j\hat \rho \hat K_j^\dagger.
\end{align}
A Kraus map is always completely positive and additionally is trace preserving if $\sum_{j=1}^n \hat K_j^\dagger \hat K_j = \hat I$.
We can expand a given map $\Lambda(\hat\rho)$ in terms of the Pauli basis by exploiting that $\sum_{j=0}^{d^2-1} \sket{j}\sbra{j} = \sum_{j=0}^{d^2-1} \sket{\hat P_j}\sbra{\hat P_j} = \hat{I}$ where $\hat{I}$ is the super-identity map.

For any given map $\Lambda(\cdot), \mathcal{B} \rightarrow \mathcal{B}$, where $\mathcal{B}$ is the space of bounded operators, we can compute its Pauli-transfer matrix as
\begin{align}
        (\mathcal{R}_\Lambda)_{jk} := \tr{\hat P_j \Lambda(\hat P_k)},\quad j,k=0,1,,\dots, d^2-1.
\end{align}
In contrast to Ref. \cite{Chow2012}, our tomography method does not rely on a measurement with continuous outcomes but rather discrete POVM outcomes $j \in \{0,1,\dots, d-1\}$,  where $d$ is the dimension of the underlying Hilbert space. In the case of perfect readout fidelity the POVM outcome $j$ coincides with a projective outcome of having measured the basis state $\ket{j}$. For imperfect measurements, we can falsely register outcomes of type $k\ne j$ even if the physical state before measurement was $\ket{j}$.  This is quantitatively captured by the readout POVM. Any detection scheme---including the actual readout and subsequent signal processing and classification step to a discrete bitstring outcome---can be characterized by its confusion rate matrix, which provides the conditional probabilities $p(j|k):= p$(detected $j$ $\mid$ prepared $k$) of detected outcome $j$ given a perfect
preparation of basis state $\ket{k}$
\begin{align}
     P = \begin{pmatrix}
            p(0 | 0)   & p(0 | 1)   & \cdots & p(0 | {d-1})  \\
            p(1 | 0)   & p(1 | 1)   & \cdots & p(1 | {d-1})  \\
            \vdots       &              &        & \vdots          \\
            p(d-1 | 0) & p(d-1 | 1) & \cdots & p(d-1 | {d-1})
        \end{pmatrix}.
\end{align}
The trace of the confusion rate matrix divided by the number of states
$F:=\tr{ P}/d = \sum_{j=0}^{d-1} p(j|j)/d$ gives the joint assignment fidelity of our simultaneous qubit readout \cite{Jeffrey2014,Magesan2014}. Given the coefficients appearing in the confusion rate matrix the equivalent readout POVM is
\begin{align}
    \hat N_j := \sum_{k=0}^{d-1} p(j | k) \hat\Pi_{k}
\end{align}
where we have introduced the bitstring projectors $\hat \Pi_{k}=\ket{k}\bra{k}$. We can immediately see that $\hat N_j\ge 0$ for all $j$, and verify the normalization
\begin{align}
    \sum_{j=0}^{d-1}\hat N_j = \sum_{k=0}^{d-1} \underbrace{\sum_{j=0}^{d-1} p(j | k)}_{1} \hat \Pi_{k}
    = \sum_{k=0}^{d-1} \hat \Pi_{k} = \mathbb{\hat I}
\end{align}
where $\mathbb{\hat I}$ is the identity operator.

\subsection{State tomography}
For state tomography, we use a control sequence to prepare a state $\rho$ and then apply $d^2$ different post-rotations $\hat R_k$ to our state $\rho \mapsto \Lambda_{R_k}(\hat \rho) := \hat R_k\hat\rho \hat R_k^\dagger$ such that $\vec{\Lambda_{R_k}(\hat \rho)} = \doublehat{\Lambda}_{R_k} \sket{\rho}$ and subsequently measure it in our given measurement basis. We assume that subsequent measurements are independent which implies that the relevant statistics for our Maximum-Likelihood-Estimator (MLE) are the histograms of measured POVM outcomes for each prepared state:
\begin{align}
        n_{jk} := \text{ number of outcomes } j \text{ for an initial state } \doublehat{\Lambda}_{R_k} \sket{\rho}
\end{align}
If we measure a total of $n_k = \sum_{j=0}^{d-1} n_{jk}$ shots for the pre-rotation $\hat R_k$ the probability of obtaining the outcome $h_k:=(n_{0k}, \dots, n_{(d-1)k})$ is given by the multinomial distribution
\begin{align}
        p(h_k) = {n_k \choose n_{0k} \; n_{1k} \; \cdots \; \; n_{(d-1)k}} p_{0k}^{n_{0k}} \cdots p_{(d-1)k}^{n_{(d-1)k}},
\end{align}
where for fixed $k$ the vector $(p_{0k},\dots, p_{(d-1)k})$ gives the single shot probability over the POVM outcomes for the prepared circuit. These probabilities are given by
\begin{align}
    \begin{split}
        p_{jk} &:= \sbra{N_j}\doublehat{\Lambda}_{R_k}\sket{\rho} \\
        &= \sum_{m=0}^{d^2-1}\underbrace{\sum_{r=0}^{d^2-1}\pi_{jr}(\mathcal{\hat R}_{k})_{rm}}_{C_{jkm}}\rho_m \\
        &= \sum_{m=0}^{d^2-1} C_{jkm}\rho_m.
    \end{split}
\end{align}
Here we have introduced $\pi_{jl}:=\sbraket{ N_j}{ P_l} = \tr{\hat N_j \hat P_l}$, $(\mathcal{R}_{k})_{rm}:= \sbra{P_r}\doublehat{\Lambda}_{R_k}\sket{P_m}$ and $\rho_m:= \sbraket{P_m}{ \rho}$. The POVM operators $ N_j = \sum_{k=0}^{d-1} p(j |k)  \Pi_{k}$ are defined as above.

The joint log likelihood for the unknown coefficients $\rho_m$ for all pre-measurement channels $\mathcal{R}_k$ is given by
\begin{align}
    \log L (\rho) = \sum_{j=0}^{d-1}\sum_{k=0}^{d^2-1} n_{jk}\log\left(\sum_{m=0}^{d^2-1} C_{jkm} \rho_m\right) + {\rm const}.
\end{align}

Maximizing this is a convex problem and can be efficiently done even with constraints that enforce normalization $\tr{\rho}=1$ and positivity $\rho \ge 0$.

\subsection{Process Tomography}
Process tomography introduces an additional index over the pre-rotations $\hat R_l$ that act on a fixed initial state $\rho_0$. The result of each such preparation is then acted on by the process $\doublehat \Lambda$ that is to be inferred. This leads to a sequence of different states 
\begin{align}
\hat \rho^{(kl)}:= \hat R_k\Lambda(\hat R_l \rho_0 \hat R_l^\dagger)\hat R_k^\dagger \leftrightarrow \sket{\rho^{(kl)}} = \doublehat{\Lambda}_{R_k} \doublehat{\Lambda} \doublehat{\Lambda}_{R_l}\sket{\rho_0}.
\end{align}
The joint histograms of all such preparations and final POVM outcomes is given by 
\begin{align}
    n_{jkl} := \text{ number of outcomes } j \text{ given input } \sket{\rho^{(kl)}}.
\end{align}
If we measure a total of $n_{kl} = \sum_{j=0}^{d-1} n_{jkl}$ shots for the post-rotation $k$
and pre-rotation $l,$ the probability of obtaining the outcome $m_{kl}:=(n_{0kl}, \dots, n_{(d-1)kl})$ is given by the binomial
\begin{align}
        p(m_{kl}) = {n_{kl} \choose n_{0kl} \; n_{1kl} \; \cdots \; \; n_{(d-1)kl}} p_{0kl}^{n_{0kl}} \cdots p_{(d-1)kl}^{n_{(d-1)kl}}
\end{align}
where the single shot probabilities $p_{jkl}$ of measuring outcome $N_j$ for the
post-channel $k$ and pre-channel $l$ are given by
\begin{align}
    \begin{split}
        p_{jkl} &:= \sbra{N_j}\doublehat{\Lambda}_{R_k} \doublehat{\Lambda} \doublehat{\Lambda}_{R_l}\sket{\rho_0} \\
        &= \sum_{m,n=0}^{d^2-1}\underbrace{\sum_{r,q=0}^{d^2-1}\pi_{jr}(\mathcal{R}_{k})_{rm} (\mathcal{R}_{l})_{nq} (\rho_0)_q}_{B_{jklmn}}(\mathcal{R})_{mn} \\
        &= \sum_{mn=0}^{d^2-1} B_{jklmn}(\mathcal{R})_{mn}
    \end{split}
\end{align}
where $\pi_{jl}:=\sbraket{N_j}{l} = \tr{\hat N_j \hat P_l}$
and $(\rho_0)_q := \sbraket{ P_q}{\rho_0} = \tr{\hat P_q \hat \rho_0}$ and the Pauli-transfer matrices for the pre and post rotations $R_l$ and the unknown process are given by
\begin{align}
        (\mathcal{R}_{l})_{nq} &:= \tr{\hat P_n \hat R_l \hat P_q \hat R_l^\dagger},\\
        \mathcal{R}_{mn} &:= \tr{\hat P_m \Lambda(\hat R_n)}.
\end{align}
The joint log likelihood for the unknown transfer matrix $\mathcal{R}$ for all pre-rotations
$\mathcal{R}_l$ and post-rotations $\mathcal{R}_k$ is given by
\begin{align}
        \log L (\mathcal{R}) = \sum_{j=0}^{d-1} \sum_{kl=0}^{d^2-1} n_{jkl}\log\left(\sum_{mn=0}^{d^2-1} B_{jklmn} (\mathcal{R})_{mn}\right) + {\rm const}.
\end{align}
Handling positivity constraints is achieved by constraining the associated Choi-matrix
to be positive \cite{Chow2012}. We can also constrain the estimated transfer matrix to preserve the trace of the mapped state by demanding that $\mathcal{R}_{0l}=\delta_{0l}$.

\subsection{Crosstalk QPT}

In the main text, we describe two-qubit QPT experiments under the preparation of 64 unique bitstrings of a separate six-qubit register. In Table~\ref{CircuitFidelity}, we present the resulting estimated average gate fidelity for each circuit.

\begin{table*}
\caption{Averaged quantum process fidelity for a different preparation states for the register of six ancilla qubits. A histogram of these results can be found in the main text.}
\label{CircuitFidelity}
\begin{tabular}{cccp{1cm}ccc}
\toprule
$N_{\mathrm{Excitations}}$ & Circuit &  Fidelity & & $N_{\mathrm{Excitations}}$ & Circuit &  Fidelity \\
0 &  $I_2I_3I_4I_5I_6I_7$ &      0.92 & & 3 &  $X_2I_3I_4I_5X_6X_7$ &      0.92 \\
1 &  $I_2I_3I_4I_5I_6X_7$ &      0.93 & & 3 &  $X_2I_3I_4X_5I_6X_7$ &      0.95 \\
1 &  $I_2I_3I_4I_5X_6I_7$ &      0.91 & & 3 &  $X_2I_3I_4X_5X_6I_7$ &      0.94 \\
1 &  $I_2I_3I_4X_5I_6I_7$ &      0.93 & & 3 &  $X_2I_3X_4I_5I_6X_7$ &      0.92 \\
1 &  $I_2I_3X_4I_5I_6I_7$ &      0.87 & & 3 &  $X_2I_3X_4I_5X_6I_7$ &      0.94 \\
1 &  $I_2X_3I_4I_5I_6I_7$ &      0.94 & & 3 &  $X_2I_3X_4X_5I_6I_7$ &      0.94 \\
1 &  $X_2I_3I_4I_5I_6I_7$ &      0.82 & & 3 &  $X_2X_3I_4I_5I_6X_7$ &      0.96 \\
2 &  $I_2I_3I_4I_5X_6X_7$ &      0.91 & & 3 &  $X_2X_3I_4I_5X_6I_7$ &      0.95 \\
2 &  $I_2I_3I_4X_5I_6X_7$ &      0.92 & & 3 &  $X_2X_3I_4X_5I_6I_7$ &      0.91 \\
2 &  $I_2I_3I_4X_5X_6I_7$ &      0.90 & & 3 &  $X_2X_3X_4I_5I_6I_7$ &      0.87 \\
2 &  $I_2I_3X_4I_5I_6X_7$ &      0.93 & & 4 &  $I_2I_3X_4X_5X_6X_7$ &      0.91 \\
2 &  $I_2I_3X_4I_5X_6I_7$ &      0.96 & & 4 &  $I_2X_3I_4X_5X_6X_7$ &      0.89 \\
2 &  $I_2I_3X_4X_5I_6I_7$ &      0.93 & & 4 &  $I_2X_3X_4I_5X_6X_7$ &      0.88 \\
2 &  $I_2X_3I_4I_5I_6X_7$ &      0.91 & & 4 &  $I_2X_3X_4X_5I_6X_7$ &      0.90 \\
2 &  $I_2X_3I_4I_5X_6I_7$ &      0.91 & & 4 &  $I_2X_3X_4X_5X_6I_7$ &      0.91 \\
2 &  $I_2X_3I_4X_5I_6I_7$ &      0.95 & & 4 &  $X_2I_3I_4X_5X_6X_7$ &      0.90 \\
2 &  $I_2X_3X_4I_5I_6I_7$ &      0.94 & & 4 &  $X_2I_3X_4I_5X_6X_7$ &      0.92 \\
2 &  $X_2I_3I_4I_5I_6X_7$ &      0.85 & & 4 &  $X_2I_3X_4X_5I_6X_7$ &      0.87 \\
2 &  $X_2I_3I_4I_5X_6I_7$ &      0.90 & & 4 &  $X_2I_3X_4X_5X_6I_7$ &      0.89 \\
2 &  $X_2I_3I_4X_5I_6I_7$ &      0.94 & & 4 &  $X_2X_3I_4I_5X_6X_7$ &      0.91 \\
2 &  $X_2I_3X_4I_5I_6I_7$ &      0.92 & & 4 &  $X_2X_3I_4X_5I_6X_7$ &      0.92 \\
2 &  $X_2X_3I_4I_5I_6I_7$ &      0.91 & & 4 &  $X_2X_3I_4X_5X_6I_7$ &      0.89 \\
3 &  $I_2I_3I_4X_5X_6X_7$ &      0.92 & & 4 &  $X_2X_3X_4I_5I_6X_7$ &      0.95 \\
3 &  $I_2I_3X_4I_5X_6X_7$ &      0.91 & & 4 &  $X_2X_3X_4I_5X_6I_7$ &      0.92 \\
3 &  $I_2I_3X_4X_5I_6X_7$ &      0.93 & & 4 &  $X_2X_3X_4X_5I_6I_7$ &      0.91 \\
3 &  $I_2I_3X_4X_5X_6I_7$ &      0.94 & & 5 &  $I_2X_3X_4X_5X_6X_7$ &      0.88 \\
3 &  $I_2X_3I_4I_5X_6X_7$ &      0.92 & & 5 &  $X_2I_3X_4X_5X_6X_7$ &      0.89 \\
3 &  $I_2X_3I_4X_5I_6X_7$ &      0.92 & & 5 &  $X_2X_3I_4X_5X_6X_7$ &      0.93 \\
3 &  $I_2X_3I_4X_5X_6I_7$ &      0.90 & & 5 &  $X_2X_3X_4I_5X_6X_7$ &      0.90 \\
3 &  $I_2X_3X_4I_5I_6X_7$ &      0.90 & & 5 &  $X_2X_3X_4X_5I_6X_7$ &      0.96 \\
3 &  $I_2X_3X_4I_5X_6I_7$ &      0.91 & & 5 &  $X_2X_3X_4X_5X_6I_7$ &      0.93 \\
3 &  $I_2X_3X_4X_5I_6I_7$ &      0.94 & & 6 &  $X_2X_3X_4X_5X_6X_7$ &      0.94 \\
\end{tabular}
\end{table*}

\end{document}